\documentclass[twocolumn]{autart}    
\usepackage{picins}
\usepackage{graphicx}
\usepackage{amssymb}
\usepackage{algorithm}
\usepackage{algpseudocode}
\usepackage{bm}
\usepackage{url}
\usepackage{color}

\def\bpf{\textnormal{\textbf{Proof.}\hspace{1ex}}}
\def\epf{\hfill \mbox{\qed}}
\newtheorem{problem}{Problem}
\newtheorem{lemma}{Lemma}
\newtheorem{remark}{Remark}
\newtheorem{definition}{Definition}
\begin{document}
\begin{frontmatter}

\title{Robust MPC for tracking of nonholonomic robots with additive disturbances\thanksref{footnoteinfo}} 

\thanks[footnoteinfo]{This paper was not presented at any IFAC
meeting. Corresponding author Y.~Xia. Tel. +86 10 68914350.
Fax +86 10 68914382.}

\author[BIT]{Zhongqi Sun}\ead{sunzhongqi12@gmail.com},    
\author[BIT]{Li Dai}\ead{daili1887@gmail.com},
\author[BIT]{Kun Liu}\ead{kunliubit@bit.edu.cn},  
\author[BIT]{Yuanqing Xia}\ead{xia\_yuanqing@bit.edu.cn},               
\author[KTH]{Karl Henrik Johansson}\ead{kallej@kth.se}
\address[BIT]{School of Automation, Beijing Institute of Technology, Beijing 100081, China}  
\address[KTH]{ACCESS Linnaeus Centre and School of Electrical Engineering,
KTH Royal Institute of Technology, SE-100 44 Stockholm, Sweden}             

\begin{keyword}                           
Robust control; Model predictive control (MPC); Nonholonomic systems; Bounded disturbances.               
\end{keyword}                             

\begin{abstract}
In this paper, two robust model predictive control (MPC) schemes are proposed for tracking control of nonholonomic systems with bounded disturbances: tube-MPC and nominal robust MPC (NRMPC). In tube-MPC, the control signal consists of a control action and a nonlinear feedback law based on the deviation of the actual states from the states of a nominal system. It renders the actual trajectory within a tube centered along the optimal trajectory of the nominal system. Recursive feasibility and input-to-state stability are established and the constraints are ensured by tightening the input domain and the terminal region. While in NRMPC, an optimal control sequence is obtained by solving an optimization problem based on the current state, and the first portion of this sequence is applied to the real system in an open-loop manner during each sampling period. The state of nominal system model is updated by the actual state at each step, which provides additional a feedback. By introducing a robust state constraint and tightening the terminal region, recursive feasibility and input-to-state stability are guaranteed. Simulation results demonstrate the effectiveness of both strategies proposed.
\end{abstract}

\end{frontmatter}
%

\section{Introduction}
Tracking control of nonholonomic systems is a fundamental motion control problem and has broad applications in many important fields such as unmanned ground vehicle navigation \cite{simanek2015evaluation}; multi-vehicle cooperative control \cite{wang2014distributed}; formation control \cite{lafferriere2005decentralized}; and so on. So far, many techniques has been developed for control of nonholonomic robots \cite{jiangdagger1997tracking,yang1999sliding,lee2001tracking,marshall2006pursuit,ghommam2010formation}. However, these techniques either ignore the mechanical constraints, or require the persistent excitation of the reference trajectory, i.e., the linear and angular velocity must not converge to zero \cite{gu2006receding}. Model predictive control (MPC) is widely used in constrained systems. By solving a finite horizon open-loop optimization problem on-line based on the current system state at each sampling instant, an optimal control sequence is obtained. The first portion of the sequence is applied to the system at each actuator update \cite{mayne2000constrained}. MPC for tracking of noholonomic systems was studied in \cite{wang2014distributed,gu2006receding,chen2009leader,sun2016receding}, where the robots were considered to be perfectly modeled. However, when the system is uncertain or perturbed, then stability and feasibility of such MPC may be lost. In the absence of constraints and uncertainties, the optimal predictive control sequence obtained by MPC is identical to that obtained by dynamic programming (DP), which provides an optimal feedback policy or sequence of control laws \cite{rawlings2009model}. Considering that feedback control is superior to open-loop control in the aspect of robustness and that DP cannot deal with the constrained systems, design methods for MPC with robust guarantees is an urgent demand for the tracking of constrained nonholonomic systems.

There are several design methods for robust MPC. One of the simplest approaches is to ignore the uncertainties and rely on the inherent robustness of deterministic MPC \cite{scokaert1995stability,marruedo2002stability}, in which an open-loop control action solved on-line is applied recursively to the system. However, the open-loop control during each sampling period may degrade the control performance even render the system unstable. Hence, feedback MPC was proposed in \cite{kothare1996robust,lee1997worst,wan2002robust,magni2003robust}, in which a sequence of feedback control laws is obtained by solving an optimization problem. The determination of a feedback policy is usually prohibitively difficult. To overcome this difficulty, it is intuitive to focus on simplifying approximations by, for instance, solving a min-max optimization problem on-line \cite{lee1997worst,wan2002robust,magni2003robust,chen1997game,limon2006input,raimondo2009min}. Min-max MPC provides a conservative robust solution for systems with bounded disturbances by considering all possible disturbances realizations. It is in most cases computationally intractable to achieve such feedback laws, since the computational complexity of min-max MPC grows exponentially with the increase of the prediction horizon.

Tube-MPC taking advantage both open-loop and feedback MPC was reported in \cite{mayne2001robustifying,chisci2001systems,langson2004robust,mayne2005robust,mayne2011tube,yu2013tube,fleming2015robust}. Here the controller consists of an optimal control action and a feedback control law. The optimal control action steers the state to the origin asymptotically, and the feedback control law maintains the actual state within a ``tube'' centered along the optimal state trajectory. Tube-MPC for linear systems was advocated in \cite{mayne2001robustifying,chisci2001systems,langson2004robust}, where the center of the tube was provided by employing a nominal system and the actual trajectory was restricted by an affine feedback law. It was shown that the computational complexity is linear rather than exponential with the increase of prediction horizon. The authors of \cite{mayne2005robust} took the initial state of the nominal system employed in the optimization problem as a decision variable in addition to the traditional control sequence, and proved several potential advantages of such an approach. Tube-MPC for  nonlinear systems with additive disturbances was studied in \cite{mayne2011tube,yu2013tube}, where the controller possessed a similar structure as in the linear case but the feedback law was replaced by another MPC to attenuate the effect of disturbances. Two optimization problems have to be solved on-line, which increases the computation burden.

In fact, tube-MPC provides a suboptimal solution because it has to tighten the input domain in the optimization problem, which may degrade the control capability. It is natural to inquire if nominal MPC is sufficiently robust to disturbances. A robust MPC via constraint restriction was developed in \cite{chisci2001systems} for regulation of discrete-time linear systems, in which asymptotic state regulation and feasibility of the optimization problem were guaranteed. In \cite{marruedo2002input}, a robust MPC for discrete-time nonlinear system using nominal predictions was presented. By tightening the state constraints and choosing a suitable terminal region, robust feasibility and input-state-stability was guaranteed. In \cite{richards2006robust}, the authors designed a constraint tightened in a monotonic sequence in the optimization problem such that the solution is feasible for all admissible disturbances. A novel robust dual-mode MPC scheme for a class of nonlinear systems was proposed in \cite{li2014robust}, the system of which is assumed to be linearizable. Since the procedure of this class of robust MPC is almost the same as nominal MPC, we call this class of robust MPC as nominal robust MPC (NRMPC) in this paper.

Robust MPC for linear systems is well studied but for nonlinear systems is still challenging since it is usually intractable to design a feedback law yielding a corresponding robust invariant set. Especially, the study of robust MPC for nonholonomic systems remains open. Motivated by the analysis above, this paper focuses on the design of robust MPC for tracking of nonholonomic systems with coupled input constraint and  bounded additive disturbances. We discuss two robust MPC schemes introduced above. First, a tube-MPC strategy with two degrees of freedom is developed, in which the nominal system is employed to generate a central trajectory and a nonlinear feedback is designed to steer the system trajectory of actual system within the tube for all admissible disturbances. Recursive feasibility and input-to-state stability are guaranteed by tightening the input domain and terminal constraint via affine transformation and all the constraints are ensured. Since tube-MPC sacrifices optimality for simplicity, an NRMPC strategy is presented, in which the state of the nominal system is updated by the actual one in each step. In such a way, the control action applied to the real system is optimal with respect to the current state. Input-to-state stability is also established by utilizing the recursive feasibility and the tightened terminal region.

The remainder of this paper is organized as follows. In Section~\ref{sec problem}, we outline the control problem and some preliminaries. Tube-MPC and NRMPC schemes are developed in Section~\ref{sec tube-MPC} and Section~\ref{sec NRMPC}, respectively, for tracking of nonholonomic systems . In Section~\ref{sec simulation}, Simulation results are given. Finally, we summarize the works of this paper in Section~\ref{sec conclusion}.

\emph{Notation}: $\mathbb{R}$ denotes the real space and $\mathbb{N}$ denotes the collection of all positive integers. For a given matrix $M$, $\|M\|$ denotes its 2-norm. $\mathrm{diag}\{x_1, x_2, \dots, x_n\}$ denotes the diagonal matrix with entries $x_1, x_2, \dots, x_n \in \mathbb{R}$. For two vectors $x = [x_1, x_2, \dots, x_n]^\mathrm{T}$ and $y = [y_1, y_2, \dots, y_n]^\mathrm{T}$, $x < y$ means $\{x_1 < y_1, x_2 < y_2, \dots, x_n < y_n\}$ and $|x|\triangleq[|x_1|, |x_2|, \dots, |x_n|]^\mathrm{T}$ denotes its absolute value. $\|x\|\triangleq\sqrt{x^{\mathrm{T}}x}$ is the Euclidean norm. $P$-weighted norm is denoted as $\|x\|_P\triangleq\sqrt{x^{\mathrm{T}}Px}$, where $P$ is a positive define matrix with appropriate dimension. Given two sets $\mathbb{A}$ and $\mathbb{B}$, $\mathbb{A}\oplus \mathbb{B} \triangleq \{a+b\left|\right.a\in \mathbb{A}, b\in \mathbb{B}\}$, $\mathbb{A}\ominus \mathbb{B} \triangleq \{a\left|\right.\{a\}\oplus \mathbb{B} \subset \mathbb{A}\}$ and $M\mathbb{A}\triangleq \{Ma | a\in \mathbb{A}\}$, where $M$ is a matrix with appropriate dimensions.

\section{Problem formulation and preliminaries} \label{sec problem}
In this section, we first introduce the kinematics of the nonholonomic robot and deduce the coupled input constraint from its mechanical model. Then, we formulate the tracking problem as our control objective, and finally give some preliminaries for facilitating the development of our main results.
\subsection{Kinematics of the nonholonomic robot}
Consider the nonholonomic robot described by the following unicycle-modeled dynamics:
\begin{equation}\label{dynamics}
  \dot{\xi}(t) = f(\xi(t), u(t))
  = \left[
    \begin{array}{c c}
      \cos\theta(t) &0\\
      \sin\theta(t) & 0\\
      0 & 1 \\
    \end{array}
  \right]u(t),
\end{equation}
where $\xi(t) = [p^\mathrm{T}(t), \theta(t)]^\mathrm{T}\in \mathbb{R}^2\times(-\pi, \pi]$ is the state, consisting of position $p(t) = [x(t), y(t)]^\mathrm{T}$ and orientation $\theta(t)$, and $u(t) = [v(t), \omega(t)]^\mathrm{T}$ is the control input with the linear velocity $v(t)$ and the angular velocity $\omega(t)$.
\begin{figure}[!htb]
  \centering
  \includegraphics[width=2.5in]{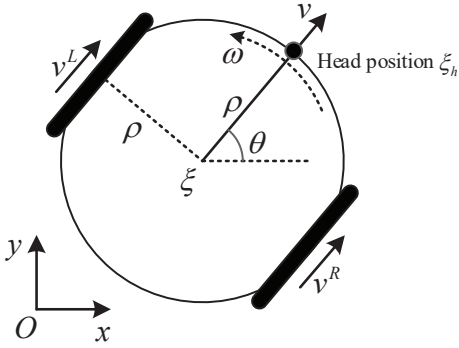}\\
  \caption{The structure of the nonholonomic robot}\label{Unicycle}
\end{figure}

The structure of the nonholonomic robot is shown in Fig.~\ref{Unicycle}. $\rho$ is half of the wheelbase, $v^L$ and $v^R$ are the velocities of the left and the right driving wheels of the robot, respectively. Denote $\xi_h$ by the head position which is the point that lies a distance $\rho$ along the perpendicular bisector of the wheel axis ahead of the robot and is given by
\begin{equation}\label{head position}
  \xi_{h}(t) =
  \left[
    \begin{array}{c}
      x_{h}(t) \\
      y_{h}(t)\\
      \theta_{h}(t)\\
    \end{array}
  \right] =
  \left[
    \begin{array}{c}
      x(t) \\
      y(t)\\
      \theta(t)\\
    \end{array}\right]+\rho
    \left[
    \begin{array}{c}
      \cos\theta(t) \\
      \sin\theta(t)\\
      0\\
    \end{array}\right].
\end{equation}
The nominal system of the head position is then formulated as
\begin{equation}\label{nominal system}
  \dot{\xi}_{h}(t)= f_{h}(\xi_{h}(t), u(t))\!=\!
  \left[
    \begin{array}{c c}
      \cos\theta(t) & -\rho\sin\theta(t)\\
      \sin\theta(t) & \rho\cos\theta(t)\\
      0 & 1 \\
    \end{array}
  \right]u(t).
\end{equation}

It is assumed that the two wheels of the robot possess the same mechanical properties and are bounded by $|v^{L}| \leq a$ and $|v^{R}| \leq a$, where $a \in \mathbb{R}$ is a known positive constant. The linear and angular velocities of the robot are presented as
\begin{eqnarray}
  v &=& (v^L + v^R)/2, \nonumber \\
  \omega &=& (v^R - v^L)/2\rho.
\end{eqnarray}
As a consequence, the control input $u$ should satisfy the constraint $u\in \mathbb{U}$, where
\begin{equation} \label{diamond constraint}
\mathbb{U} = \{[v, \omega]^\mathrm{T}: \frac{|v|}{a} +\frac{|\omega|}{b} \leq 1\}
\end{equation}
with $b = a/\rho$.

\subsection{Control objective}

Our control objective is to track a reference trajectory in a global frame $O$. The reference trajectory, which can be viewed as a virtual leader, is described by a reference state vector $\xi_r(t)
 = [p_r^\mathrm{T}(t), \theta_r(t)]^\mathrm{T} \in \mathbb{R}^2\times(-\pi, \pi]$ with $p_r(t) = [x_r(t), y_r(t)]^\mathrm{T}$ and a reference control signal $u_r(t) = [v_r(t), \omega_r(t)] \in \mathbb{U}$. The reference state vector $\xi_r(t)$ and the reference control signal $u_r(t)$ are modeled as a nominal unicycle robot
\begin{equation}\label{reference}
  \dot{\xi_r}(t) = f(\xi_r(t), u_r(t)).
\end{equation}

The follower to be controlled is also an unicycle with kinematics (\ref{dynamics}). Considering the existence of nonholonomic constraint, we consider its head position modeled as (\ref{nominal system}). Furthermore, the robot is assumed to be perturbed by a disturbance caused by sideslip due to the road ride. Therefore, we consider disturbances acting on the linear velocity while neglecting disturbances acting on the angular velocity. The perturbed head position kinematics is then formulated as follows:
\begin{equation}\label{disturbed follower system}
  \dot{\xi}_{fh}(t) = f_{h}(\xi_{fh}(t), u_f(t)) + d(t), \quad u_f(t) \in \mathbb{U},
\end{equation}
where $\xi_{fh}(t) = [p_{fh}^\mathrm{T}(t), \theta_{f}(t)]^\mathrm{T}$ is the state with the head position $p_{fh}(t) =[x_{fh}(t), y_{fh}(t)]^\mathrm{T}$, $u_f(t)=[v_f(t), \omega_f(t)]^\mathrm{T}$ is the control input, and $d(t)=[d_p^\mathrm{T}(t), 0]^\mathrm{T} \subseteq \mathbb{R}^3$, $d_p(t) = [d_x(t), d_y(t)]^\mathrm{T}$, is the external disturbances, which is bounded by $\|d(t)\|\leq \eta$.

\begin{figure}[htb]
  \centering
  \includegraphics[width=2.8in]{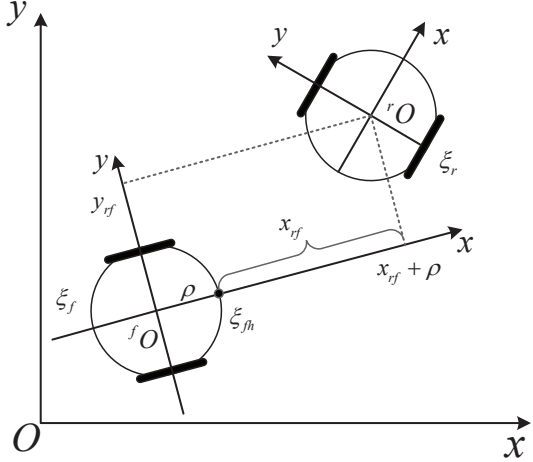}\\
  \caption{Leader-follower configuration}\label{formation}
\end{figure}

Construct Frenet-Serret frames $^rO$ and $^fO$ for the virtual leader and the follower, respectively. They are moving frames fixed on the robots (see Fig.~\ref{formation}).
The tracking error $p_{rf} = [x_{rf}, y_{rf}]^\mathrm{T}$ with respect to the Frenet-Serret frame $^fO$ is given by
\begin{eqnarray}\label{tracking error}
p_{rf}(t) &=& R(\theta_f(t))(p_r(t) - p_f(t)),\\
\theta_{rf}(t) &=& \theta_r(t) - \theta_f(t),
\end{eqnarray}
where $R(\theta) = \left[
                       \begin{array}{cc}
                         \cos\theta & -\sin\theta \\
                         \sin\theta &  \cos\theta\\
                       \end{array}
                     \right]
$ is the rotation matrix.

Taking the derivative of the tracking error yields
\begin{eqnarray}\label{error dynamics}
  &&\dot{p}_{rf}(t) =
     \left[
               \begin{array}{cc}
                 0 & \omega_f(t) \\
                 -\omega_f(t) & 0 \\
               \end{array}
             \right]
             \left[
               \begin{array}{c}
                 x_{rf}(t) \\
                 y_{rf}(t) \\
               \end{array}
             \right]\nonumber\\
         && +
     \left[\begin{array}{c}
     -v_f(t) +  v_r(t) \cos\theta_{rf}(t) \\
  -\rho\omega_f(t) + v_r(t) \sin\theta_{rf}(t)
     \end{array}\right]
     +
     R(\theta_f)
     \left[
       \begin{array}{c}
         d_x(t) \\
         d_y(t) \\
       \end{array}
     \right].
\end{eqnarray}

Based on the discussion above, we will design robust MPC strategies to drive the tracking error $p_{rf}$ to a neighborhood of the origin. Note that the tracking system (\ref{error dynamics}) involves the disturbances but the future disturbances cannot be predicted in advance. We will formulate the MPC problem only involving the nominal system.

%
%
To distinguish the variables in the nominal system model from the real system, we introduce $\tilde{\cdot}$ as a superscript for the variables in the nominal system. From the perturbed system~(\ref{disturbed follower system}), the nominal dynamics can be obtained by neglecting the disturbances as
\begin{equation}\label{nominal follower system}
  \dot{\tilde{\xi}}_{fh}(t) = f_{h}(\tilde{\xi}_{fh}(t), \tilde{u}_f(t)), \quad \tilde{u}_f(t) \in \mathbb{U},
\end{equation}
where, similarly, $\tilde{\xi}_{fh}(t) = [\tilde{p}_{fh}^\mathrm{T}(t), \tilde{\theta}_{f}(t)]^\mathrm{T}$ is the state of the nominal system with the position $\tilde{p}_{fh} =[\tilde{x}_{fh}(t), \tilde{y}_{fh}(t)]^\mathrm{T}$ and orientation $\tilde{\theta}_{f}(t)$, and $\tilde{u}_f(t) = [\tilde{v}_f(t), \tilde{\omega}_f(t)]^\mathrm{T}$ is the control input of the nominal system. The tracking error dynamics based on the nominal system is then given by
\begin{equation}\label{nominal tracking error}
 \dot{\tilde{p}}_{rf}(t) = 
     \left[
               \begin{array}{cc}
                 0 & \omega_f(t) \\
                 -\omega_f(t) & 0 \\
               \end{array}
             \right]
             \left[
               \begin{array}{c}
                 \tilde{x}_{rf}(t) \\
                 \tilde{y}_{rf}(t) \\
               \end{array}
             \right]\\
             +\tilde{u}_{rf}(t),
\end{equation}
where $\tilde{u}_{rf}(t)$ is the input error and is given by
\begin{equation}\label{}
  \tilde{u}_{rf}(t) =
  \left[\begin{array}{c}
     -v_f(t) +  v_r(t) \cos\tilde{\theta}_{rf}(t) \\
  -\rho\omega_f(t) + v_r \sin\tilde{\theta}_{rf}(t)
     \end{array}\right].
\end{equation}

Define $\{t_k: k\in \mathbb{N}, t_{k+1}-t_k=\delta\}$, with $\delta>0$, the time sequence at which the open-loop optimization problems are solved. The MPC cost to be minimized is given by
\begin{eqnarray}\label{cost}
 J(\tilde{p}_{rf}(t_k), \tilde{u}_{rf}(t_k)) =&& \int_{t_k}^{t_k+T} L(\tilde{p}_{rf}(\tau|t_k), \tilde{u}_{rf}(\tau|t_k)) d\tau \nonumber\\
       && + g(\tilde{p}_{rf}(t_k+T|t_k)),
\end{eqnarray}
in which
$L(\tilde{p}_{rf}(\tau|t_k), \tilde{u}_{rf}(\tau|t_k)) = \|\tilde{p}_{rf}(\tau|t_k)\|_Q^2+\|\tilde{u}_{rf} (\tau|t_k)\|_P^2$ represents the stage cost with the positive define matrices $P = \mathrm{diag}\{p_1, p_2\}$ and $Q = \mathrm{diag}\{q_1, q_2\}$, $g(\tilde{p}_{rf}(\tau|t_k)) = \frac{1}{2}\|\tilde{p}_{rf}(\tau|t_k)\|^2$
is the terminal penalty, and $T$ is the prediction horizon satisfying $T=N\delta$, $N\in \mathbb{N}$.

\subsection{Preliminaries}
Some definitions and lemmas used in the following sections are summarized as follows.
\begin{definition}\label{def terminal state region}
For the nominal tracking error system (\ref{nominal tracking error}), the terminal region $\Omega$ and the terminal controller $\tilde{u}_f^\kappa(\cdot)$ are such that if $\tilde{p}_{rf}(t_k + T|t_k)\in\Omega$, then, for any $\tau \in (t_k + T, t_{k+1}+T]$, by implementing the terminal controller $\tilde{u}_f(\tau|t_{k+1}) = \tilde{u}_f^\kappa(\tau|t_{k+1})$, it holds that
 \begin{eqnarray}
  &&\tilde{p}_{rf}(\tau|t_k) \in\Omega, \\
  &&\tilde{u}_f(\tau|t_k) \in \mathbb{U}, \\
  &&\dot{g}(\tilde{p}_{rf}(\tau|t_k)) +  L(\tilde{p}_{rf}(\tau|t_k), \tilde{u}_{rf}(\tau|t_k)) \leq 0. \label{stability condition}
\end{eqnarray}
\end{definition}

\begin{definition}\label{ISS}(\cite{sontag1995characterizations})
System (\ref{error dynamics}) is input-to-state stable (ISS) if there exist a $\mathcal{KL}$ function $\beta(\cdot,\cdot): \mathbb{R}_{\geq0}\times \mathbb{R}_{\geq0}\rightarrow \mathbb{R}$ and a $\mathcal{K}$ function $\gamma(\cdot)$ such that, for $t\geq 0$, it holds that
\begin{equation}\label{}
  \|p_{rf}(t)\|\leq\beta(\|{p}_{rf}(t_0)\|, t) + \gamma(\eta).
\end{equation}
\end{definition}
\begin{definition}\label{ISS-Lyapunov}(\cite{rawlings2009model})
A function $V(\cdot)$ is called an ISS-Lyapunov function for system (\ref{error dynamics}) if there exist $\mathcal{K}_\infty$ functions $\alpha_1(\cdot)$, $\alpha_2(\cdot)$, $\alpha_3(\cdot)$ and a $\mathcal{K}$ function $\sigma(\cdot)$ such that for all $p_{rf} \in \mathbb{R}^2$
  \begin{eqnarray}\label{}
  &&\!\!  \alpha_1(\|p_{rf}(t_k)\|)\leq V(p_{rf}(t_k)) \leq \alpha_2(\|p_{rf}(t_k)\|),\\
  &&\!\! V(p_{rf}(t_{k+1})) \!-\! V(p_{rf}(t_{k})) \!\leq\!
    -\alpha_3(\|p_{rf}(t_{k})\|) \!+\! \sigma(\eta).
  \end{eqnarray}
\end{definition}

\begin{remark}
It should be mentioned that both Definition~\ref{ISS} and Definition~\ref{ISS-Lyapunov} result in input-to-state stability, which implies that the tracking error vanishes if there is no disturbance.
\end{remark}

The following lemma provides a terminal controller and the corresponding terminal region for the nominal error system (\ref{nominal tracking error}).
\begin{lemma}\label{lemma terminal region}
For the nominal tracking system (\ref{nominal tracking error}), let $\tilde{u}_f \in \lambda_f\mathbb{U}$ with $\lambda_f\in(0, 1]$, $|v_r| < \frac{a}{\sqrt{2}}\lambda_f$, and $\lambda_r = \frac{\sqrt{2}}{a}\max |v_r|$. Then $\Omega =  \{\tilde{p}_{rf}: \tilde{k}_1 |\tilde{x}_{rf}|+\tilde{k}_2 |\tilde{y}_{rf}| < a(\lambda_f-\lambda_r)\}$ is a terminal region for the controller
\begin{eqnarray}\label{terminal controller}
  \tilde{u}_f^\kappa(\tau|t_k)=
   \left[
     \begin{array}{c}
       \tilde{v}_f^\kappa(\tau|t_k) \\
       \tilde{\omega}_f^\kappa(\tau|t_k) \\
     \end{array}
   \right]
    =
 \left[
    \begin{array}{c}
      \tilde{k}_1 \tilde{x}_{rf} + v_r\cos\tilde{\theta}_{rf} \\
      \frac{1}{\rho}(\tilde{k}_2 \tilde{y}_{rf}  +v_r \sin\tilde{\theta}_{rf}) \\
    \end{array}\right]\nonumber\\
    \tau \in (t_k + T, t_{k+1}+T],
\end{eqnarray}
with the parameters satisfying $p_iq_i < \frac{1}{4}$ and $\tilde{k}_i \in \left(\frac{1-\sqrt{1-4p_iq_i}}{2p_i}, \frac{1+\sqrt{1-4p_iq_i}}{2p_i}\right)$, $i = 1, 2$.
\end{lemma}
\bpf
First, consider the terminal controller
\begin{eqnarray*}
 \displaystyle\frac{|\tilde{v}_f^\kappa|}{a} &+& \frac{|\tilde{\omega}_f^\kappa|}{b}\!=\!\frac{|\tilde{k}_1 \tilde{x}_{rf}\! + \!v_r\cos\tilde{\theta}_{rf})|}{a} + \frac{|\tilde{k}_2 \tilde{y}_{rf}  +v_r \sin\tilde{\theta}_{rf}|}{a}\nonumber\\
 &\leq& \frac{1}{a}(\tilde{k}_1|\tilde{x}_{rf}|+\tilde{k}_2|\tilde{y}_{rf}|+|v_r\cos\tilde{\theta}_{rf}|+|v_r \sin\tilde{\theta}_{rf}|)\nonumber\\
 &\leq&  \lambda_f-\lambda_r+\frac{\sqrt{2}}{a}|v_r|\leq\lambda_f,
\end{eqnarray*}
which implies $u_f^\kappa\in\lambda_f\mathbb{U}$ if $\tilde{p}_{rf} \in \Omega$.

Next, choose $g(\tilde{p}_{rf}(\tau|t_k))$ as Lyapunov function. The derivative of $g(\tilde{p}_{rf}(\tau|t_k))$ with respect to $\tau$ yields
\begin{eqnarray*}
 \dot{g}(\tilde{p}_{rf}(\tau|t_k))&=& -(\tilde{k}_1{\tilde{x}_{rf}}^2(\tau|t_{k})+\tilde{k}_2{\tilde{y}_{rf}}^2(\tau|t_{k})),
\end{eqnarray*}
which means that $\Omega$ is invariant by implementing the terminal controller, i.e., $\tilde{p}_{rf}(\tau|t_k) \in \Omega$ holds for all $\tau > t_k$ once $\tilde{p}_{rf}(t_k|t_k) \in \Omega$.

Finally, for $\tilde{p}_{rf}(\tau|t_k) \in \Omega$, it follows that
\begin{eqnarray}
  &&\dot{g}(\tilde{p}_{rf}(\tau|t_k)) +  L(\tilde{p}_{rf}(\tau|t_k), \tilde{u}_{rf}(\tau|t_k)) \nonumber\\
  &=&\tilde{x}_f^e \dot{\tilde{x}}_{rf} + \tilde{y}_{rf} \dot{\tilde{y}}_{rf} + q_1\tilde{x}_{rf}^2 + q_2\tilde{y}_{rf}^2+ p_1\tilde{v}_{rf}^2 + p_2\tilde{\omega}_{rf}^2\nonumber\\
  &=&-\tilde{k}_1 \tilde{x}_{rf}^2 -  \tilde{k}_2\tilde{y}_{rf}^2 + q_1\tilde{x}_{rf}^2 + q_2\tilde{y}_{rf}^2+ p_1\tilde{v}_{rf}^2 + p_2\tilde{\omega}_{rf}^2\nonumber\\
   &=& (p_1 \tilde{k}_1^2 - \tilde{k}_1 + q_1) \tilde{x}_{rf}^2 + (p_2 \tilde{k}_2^2 - \tilde{k}_2 + q_2)\tilde{y}_{rf}^2.
\end{eqnarray}
Since $p_iq_i < \frac{1}{4}$ and $\quad \tilde{k}_i \in \left(\frac{1-\sqrt{1-4p_iq_i}}{2p_i}, \frac{1+\sqrt{1-4p_iq_i}}{2p_i}\right)$, $i = 1, 2$, the inequality $\dot{g} + L < 0$ holds.

Hence, from Definition~\ref{def terminal state region}, $\Omega$ is a terminal region associated with the terminal controller $\tilde{u}_f^\kappa(\tau|t_k)$.
\epf

The nominal system (\ref{nominal system}) is Lipschitz continuous and a corresponding Lipschitz constant is given by the following lemma.
\begin{lemma}\label{lemma Lipschitz}
System (\ref{nominal system}) with ${u} \in \mathbb{U}$ is locally Lipschitz in ${\xi}_h$ with Lipschitz constant $a$, where $a$ is the max wheel speed.
\end{lemma}
\bpf
Considering the function values of $f_{h}({\xi}_h, {u})$ at ${\xi}_{h1}$ and ${\xi}_{h2}$ with the same ${u}$, we have
\begin{eqnarray*}
  &&\|f_{h}({\xi}_{h1}, {u}) - f_{h}({\xi}_{h2}, {u})\|^2 \nonumber\\
  &=& \left\|\left[
                                                              \begin{array}{c}
                                                                {v}(\cos{\theta}_{1}-\cos{\theta}_{2}) + \rho{\omega}(\sin{\theta}_{2}-\sin{\theta}_{1}) \\
                                                                {v}(\sin{\theta}_{1}-\sin{\theta}_{2})+ \rho{\omega}(\cos{\theta}_{1}-\cos{\theta}_{2})\\
                                                                0 \\
                                                              \end{array}
                                                            \right]\right\|^2 \nonumber\\
   &=& {v}^2(\cos{\theta}_1-\cos{\theta}_2)^2 + \rho^2{\omega}^2(\sin{\theta}_{2}-\sin{\theta}_{1})^2 \nonumber\\
   & & + {v}^2(\sin{\theta}_1-\sin{\theta}_2)^2 + \rho^2{\omega}^2(\cos{\theta}_{2}-\cos{\theta}_{1})^2 \nonumber\\
   &\leq& 2({v}^2 + \rho^2{\omega}^2)({\theta}_1 - {\theta}_2)^2\nonumber\\
   &\leq& 2\max_{[{v}, {\omega}]^\mathrm{T} \in \mathbb{U}}\{{v}^2 + \rho^2{\omega}^2\}({\theta}_1 - {\theta}_2)^2 \nonumber\\
   &=& a^2({\theta}_1 - {\theta}_2)^2,
\end{eqnarray*}
where the mean value theorem and Lagrange multiplier method are used in the last inequality. The maximum of ${v}^2 + \rho^2{\omega}^2$, subject to ${|{v}|}/{a} +{|{\omega}|}/{b} \leq 1$, can be obtained by setting ${v} = \frac{a}{2}$ and ${\omega} = \frac{b}{2}$. From the results above, we conclude that
\begin{equation}\label{follower Lipschitz}
  \|f_{h}({\xi}_{h1}, {u}) - f_{h}({\xi}_{h2}, {u})\| \leq a\|{\xi}_{h1}- {\xi}_{h2}\|.
\end{equation}
\epf
\section{Tube-MPC}\label{sec tube-MPC}

In this section, a tube-MPC policy is developed, which consists of an optimal control action obtained by solving an optimization problem and a feedback law based on the deviation of the actual state from the nominal one. The controller forces the system state to stay within a tube around a sensible central trajectory. The cental trajectory is determined by the following optimization problem.

\begin{problem}\label{OP1}
\begin{eqnarray}
  \min_{\tilde{u}_f(\tau|t_k)} & & J(\tilde{p}_{rf}(t_k), \tilde{u}_{rf}(t_k)), \\
  s.t. & &\tilde{\xi}_{fh}(t_k|t_k) = \tilde{\xi}_{fh}(t_k), \label{op1 initial stage}\\
  & &\dot{\tilde{\xi}}_{fh}(\tau|t_k) = f_{h}(\tilde{\xi}_{fh}(\tau|t_k), \tilde{u}_f(\tau|t_k)), \\
   & & \tilde{u}_f(\tau|t_k) \in \mathbb{U}_{tube},  \\
   & & \tilde{p}_{rf}(t_k+T|t_k) \in \Omega_{tube}, \label{s.t.}
\end{eqnarray}
where $\mathbb{U}_{tube}= \{[\tilde{v}_f, \tilde{\omega}_f]^\mathrm{T}: \frac{|\tilde{v}_f|}{a} +\frac{|\tilde{\omega}_f|}{b} \leq \lambda_{tube}\}$ with $\lambda_{tube} = \frac{\sqrt{2}}{2} - \frac{\eta\sqrt{2}}{a}$, and $\Omega_{tube} = \{\tilde{p}_{rf}: \tilde{k}_1 |\tilde{x}_{rf}|+\tilde{k}_2 |\tilde{y}_{rf}| < a(\lambda_{tube}-\lambda_r)\}$.
\end{problem}

Solution of Problem \ref{OP1} yields the minimizing control sequence for the nominal follower system over the interval $[t_k, t_{k+T}]$:
\begin{eqnarray}\label{optimal control seq}
  \bm{\tilde{u}}_f^*(t_k) = \{\tilde{u}_f^*(t_k|t_k), \tilde{u}_f^*(t_{k+1}|t_k), \dots, \tilde{u}_f^*(t_{k+N}|t_k)\},
\end{eqnarray}
as well as the corresponding optimal trajectory:
\begin{eqnarray}\label{optimal trajectory seq}
 &&\bm{\tilde{\xi}}_{fh}^*(t_k) = \{[\tilde{p}_{fh}^*(t_k|t_k), \tilde{\theta}_f^*(t_{k}|t_k)]^\mathrm{T}, [\tilde{p}_{fh}^*(t_{k+1}|t_k),\nonumber\\
 && \tilde{\theta}_f^*(t_{k+1}|t_k)]^\mathrm{T},\dots,
   [\tilde{p}_{fh}^*(t_{k+N}|t_k), \tilde{\theta}_f^*(t_{k+N}|t_k)]^\mathrm{T}\}.
\end{eqnarray}

The robust controller for the follower over the interval $[t_k, t_{k+1}]$ is designed as
\begin{eqnarray}\label{tube controller}
  u_f(t_k) = && M^{-1}(\theta_f(t_k))M(\tilde{\theta}_f^*(t_k|t_k))\tilde{u}_f^*(t_k|t_k)\nonumber\\
  &&+ M^{-1}(\theta_f(t_k))K(p_{fh}(t_k) - \tilde{p}_{fh}^*(t_k|t_k)),
\end{eqnarray}
where $M(\theta) = \left[
                    \begin{array}{cc}
                      \cos\theta &-\rho\sin\theta \\
                      \sin\theta & \rho\cos\theta \\
                    \end{array}
                  \right]$, $K = \mathrm{diag}\{k_x, k_y\}$, $k_x < 0$, $k_y < 0$, is the feedback gain, $\tilde{u}_f^*(t_k|t_k)$ is the first control action of the optimal control sequence, and $\tilde{p}_{fh}^*(t_k|t_k))$ and $\tilde{\theta}_f^*(t_k|t_k)$ are the first portion of the optimal position and orientation, respectively.

Based on this control strategy, the procedure of tube-MPC is summarized in Algorithm \ref{algorithm1}.

\begin{algorithm}
\caption{Tube-MPC}\label{algorithm1}
\begin{algorithmic}[1]
\State At time $t_0$, initialize the nominal system state by the actual state $\tilde{\xi}_{fh}(0) = \xi_{fh}(0)$.
\State At time $t_k$, solve Problem \ref{OP1} based on nominal system to obtain the optimal control sequence $\bm{\tilde{u}}_f^*(t_k) = \arg\min_{u_f(\tau|t_k)}J_f(t_k, \tilde{p}_{rf}, \tilde{u}_{rf})$.
\State Calculate the actual control signal for the real system $ u_f(t_k) =  M^{-1}(\theta_f)[M(\tilde{\theta}_f^*)\tilde{u}_f^*(t_k|t_k)
                  +K(p_{fh}(t_k) - p_{fh}^*(t_k)]$.
\State Apply the first portion of the sequence, i.e., $\tilde{u}_f^*(t_k|t_k)$, to the nominal system, and apply $u_f(t_k)$ to the real system during the sampling interval $[t_k, t_{k+1}]$.
\State Update the state of the nominal system with $\tilde{\xi}_{fh}^*(t_{k+1})$ and the state of the real system with ${\xi}_{fh}(t_{k+1})$.
\State Update the time instant $t_k = t_{k+1}$ and go to step 2.
\end{algorithmic}
\end{algorithm}

\begin{remark}
Since the optimization problem is solved on-line at each step and the first optimal control action is employed to generate the control policy together with the feedback law, the computational complexity is determined by the nominal system. Hence, the scheme has the same computational complexity as the deterministic MPC.
\end{remark}

\begin{remark}
Due to the nonlinearity and nonholonomic constraint of the system, the optimal control action and the feedback law are combined in a different manner compared to linear systems \cite{mayne2001robustifying,chisci2001systems,langson2004robust}. This increases the difficulty of determining the tightened input constraint set $\mathbb{U}_{tube}$ such that $u_f \in \mathbb{U}$ holds. The scheme is also different from the existing works on nonlinear systems as in \cite{mayne2011tube} and \cite{yu2013tube}, in which our feedback law determined off-line is replaced with an online computation of another MPC. Hence, two optimization problems have to be solved in each step, which increases the computational burden.
\end{remark}

\begin{remark}\label{remark tube off-line}
From Algorithm \ref{algorithm1}, it can be observed that the optimization problem employs only the nominal system and thus the predictive optimal trajectory is independent of the actual state except for the initial one. From this point, the central trajectory of the tube can be calculated in a parallel or even off-line way if the initial state is known a priori. In such a way, only one feedback law is required to be calculated on-line, which reduces the on-line computational burden even further.
\end{remark}

Before stating the main results of tube-MPC, the following lemma is given to show that the feedback law renders the difference between the minimizing trajectory and the actual trajectory bounded while guaranteeing the satisfaction of the input constraint.
\begin{lemma}\label{lemma tube}
For the tracking control system (\ref{error dynamics}) with controller (\ref{tube controller}), it follows that
 \begin{enumerate}
  \item[(\romannumeral1)]the state of the real system lies in the tube $\mathbb{T} =  p_{fh}^* \oplus \mathbb{P}_{fe}$, where $\mathbb{P}_{fe} = \{p_{fe}(t): |p_{fe}|\leq\left[
            \begin{array}{c}
              - \frac{\eta}{k_1} \\
             - \frac{\eta}{k_2}  \\ \end{array}
          \right]\}$;
  \item[(\romannumeral2)] the input constraint is satisfied, i.e., $u_f \in \mathbb{U}$.
  \end{enumerate}
\end{lemma}
\bpf
Denote the deviation of the actual trajectory from the optimal trajectory as
\begin{equation}\label{deviation of actual from optimal}
  p_{fe}(t) = p_{fh}(t) - \tilde{p}_{fh}^*(t).
\end{equation}
Taking the derivative of (\ref{deviation of actual from optimal}) yields
\begin{eqnarray}\label{self error dynamics}
  \dot{p}_{fe}(t)\!=&&\!  \left[
                    \begin{array}{cc}
                      \cos\theta_f(t) & -\rho\sin\theta_f(t) \\
                      \sin\theta_f(t) & \rho\cos\theta_f(t) \\
                    \end{array}
                  \right]u_f(t)\nonumber\\
                 && -
                  \left[
                    \begin{array}{cc}
                      \cos\tilde{\theta}_f^*(t) & -\rho\sin\tilde{\theta}_f^*(t) \\
                      \sin\tilde{\theta}_f^*(t) & \rho\cos\tilde{\theta}_f^*(t) \\
                    \end{array}
                  \right]\tilde{u}_f^*(t)+d_p(t).
\end{eqnarray}

Substituting (\ref{tube controller}) into (\ref{self error dynamics}), we can conclude that
\begin{equation}\label{}
\dot{p}_{fe}(t) = K p_{fe}(t) + d_p(t),
\end{equation}
of which the solution is given by
\begin{eqnarray}
  p_{fe}(t) &=& e^{Kt}p_{fe}(0) + \int_0^t e^{K(t-\tau)}d_p(\tau)d\tau.
\end{eqnarray}
By the initialization stage (\ref{op1 initial stage}) and the upper-bound of the disturbances, it follows that
\begin{equation}\label{}
   |p_{fe}(t)|\leq 
\eta\left[
            \begin{array}{c}
              \frac{1}{k_1}e^{k_1t} - \frac{1}{k_1} \\
              \frac{1}{k_2}e^{k_2t} - \frac{1}{k_2}  \\
            \end{array}
          \right].
\end{equation}
Consequently, $p_{fe}(t) \in \mathbb{P}_{fe}(t)$, where the set $\mathbb{P}_{fe}(t)$ is defined by
\begin{equation}\label{}
  \mathbb{P}_{fe}(t) = \left\{p_{fe}(t): |p_{fe}(t)|\leq\eta\left[
            \begin{array}{c}
              \frac{1}{k_1}e^{k_1t} - \frac{1}{k_1} \\
              \frac{1}{k_2}e^{k_2t} - \frac{1}{k_2}  \\
            \end{array}
          \right]\right\}.
\end{equation}
We further define $\mathbb{P}_{fe}$ as
\begin{equation}\label{}
  \mathbb{P}_{fe} = \lim_{t\rightarrow\infty}\mathbb{P}_{fe}(t)=\left\{p_{fe}(t): |p_{fe}|\leq\left[
            \begin{array}{c}
              - \frac{\eta}{k_1} \\
             - \frac{\eta}{k_2}  \\
            \end{array}
          \right]\right\}.
\end{equation}
From (\ref{deviation of actual from optimal}) and $p_{fe} \in \mathbb{P}_{fe}$, we have
\begin{equation}\label{tube}
p_{fh} \in p_{fh}^*\oplus \mathbb{P}_{fe},
\end{equation}
i.e., the trajectory lies in the tube $\mathbb{T}$.

For (\romannumeral2), redefine the control input as
\begin{eqnarray}\label{afine input}
  u_f^a &=& M(\theta_f)u_f, \\
  u_f^{a*} &=& M(\theta_f^*)u_f^*.
\end{eqnarray}
It can be observed that $M(\cdot)$ is an affine transformation, which is equivalent to scaling $\omega_f$ ($\omega_f^*$) by $\rho$ and rotate $u_f$ ($u_f^*$) by $\theta$. Thus, to prove $u_f \in \mathbb{U}$ if $u_f^* \in \mathbb{U}_{tube}$ is equivalent to show $u_f^a \in \mathbb{U}^a$ if $u_f^{a*} \in \mathbb{U}_{tube}^a$ for every admissible $\theta_f$ and $\theta_f^*$. The sets $\mathbb{U}^a$ and $\mathbb{U}_{tube}^a$ are defined as follows:
\begin{eqnarray}
  &&\mathbb{U}^a = M(\theta_f)\{[v, \omega]^\mathrm{T}: \frac{|v|}{a} +\frac{|\omega|}{a} \leq 1\}, \\
  &&\mathbb{U}_{tube}^a = M(\theta_f^*)\{[v, \omega]^\mathrm{T}: \frac{|v|}{a} +\frac{|\omega|}{a} \leq \lambda_{tube}\}.
\end{eqnarray}

Substituting (\ref{afine input}) into (\ref{tube controller}) yields
\begin{equation}\label{}
  u_f^a = u_f^{a*} + Kp_{fe}.
\end{equation}
It is obvious that
\begin{equation}\label{}
\bigcap_{\theta_f\in(-\pi, \pi]} \mathbb{U}^a = \{[v, \omega]^\mathrm{T}:\|[v, \omega]^\mathrm{T}\|\leq a\frac{\sqrt{2}}{2}\},
\end{equation}
\begin{equation}\label{}
  \bigcup_{\theta_f^*\in(-\pi, \pi]}\mathbb{U}_{tube}^a = \{[v, \omega]^\mathrm{T}:\|[v, \omega]^\mathrm{T}\|\leq a\lambda_{tube}\},
\end{equation}
\begin{eqnarray}\label{}
 K\mathbb{P}_{fe} &=& \left\{Kp_{fe}(t): |Kp_{fe}|\leq\left[
            \begin{array}{c}
              \eta \\
             \eta  \\
            \end{array}
          \right]\right\}\nonumber\\
          &\subset& \left\{Kp_{fe}(t): \|Kp_{fe}\|\leq\sqrt{2}\eta\right\}.
\end{eqnarray}
Thus, it can be obtained that
\begin{equation}\label{}
  \bigcup_{\theta_f^*\in(-\pi, \pi]}\mathbb{U}_{tube}^a \oplus K\mathbb{P}_{fe} \subset \bigcap_{\theta_f\in(-\pi, \pi]} \mathbb{U}^a,
\end{equation}
which implies that $u_f^a \in \mathbb{U}^a$ holds for every admissible $\theta_f$ and $\theta_f^*$, and $u_f \in \mathbb{U}$ naturally holds.
\epf

\begin{remark}
Note that the input domain is independent of the feedback gain $K$, which differs from the results of linear systems in \cite{mayne2001robustifying,chisci2001systems,langson2004robust}. Meanwhile, from (\romannumeral1) in Lemma~\ref{lemma tube}, increasing $K$ will reduce the difference between the actual trajectory and the optimal one, and consequently reduce the size of the tube $\mathbb{T}$. It indicates that the steady tracking performance could be enhanced by tuning $K$.
\end{remark}
The main results of tube-MPC are given in the following theorem.
\begin{thm} \label{theorem tube}
For the tracking control system (\ref{error dynamics}) under Algorithm \ref{algorithm1}, if Problem \ref{OP1} is feasible at time $t_0$, then,
\begin{enumerate}
  \item[(\romannumeral1)] Problem \ref{OP1} is feasible for all $t > t_0$;
  \item[(\romannumeral2)] the tracking control system (\ref{error dynamics}) is ISS.
\end{enumerate}
\end{thm}
\bpf
From Lemma~\ref{lemma terminal region}, $\Omega_{tube}$ is a terminal region by letting $\lambda_f = \lambda_{tube}$. We assume that a feasible solution exists and an optimal solution $\tilde{u}_f^*(t_k)$ is found at the sampling instant $t_k$. When applying this sequence to the nominal system, the tracking error of the nominal system is driven into the terminal region $\Omega_{tube}$, i.e., $\tilde{p}_f^{e*}(t_k+T|t_k) \in \Omega_{tube}$, along the corresponding open-loop trajectory $\tilde{\xi}_f^*(t_k)$ over $[t_k, t_k+T]$. In terms of Algorithm \ref{algorithm1}, the open-loop control $\tilde{u}^*(t_k|t_k)$ is applied to the nominal system, and its state measurement at time $t_{k+1}$ is given by $\tilde{\xi}(t_{k+1}) = \tilde{\xi}^*(t_{k+1}|t_k)$. Therefore, to solve the open-loop optimal control problem at $t_{k+1}$ with the initial condition, a feasible solution can be constructed by
\begin{equation}\label{feasible input 1}
      \tilde{u}_f(\tau|t_{k+1}) =
      \left\{
      \begin{array}{ll}
       \tilde{u}_f^*(\tau|t_{k}), & \hbox{$\tau \in [t_{k+1}, t_k+T)$,} \\
       \tilde{u}_f^\kappa(\tau|t_k), & \hbox{$\tau \in [t_{k}+T, t_{k+1}+T$)},
     \end{array}
      \right.
     \end{equation}
 where $\tilde{u}_f^\kappa(\tau|t_k)$ is the terminal controller given by (\ref{terminal controller}). Since the terminal region $\Omega_{tube}$ is invariant with the control $\tilde{u}_f^\kappa(\tau|t_k)$, $\tilde{p}_f^{e*}(t_k+T|t_k) \in \Omega_{tube}$ implies $\tilde{p}_f^{e}(t_{k+1}+T|t_{k+1}) \in \Omega_{tube}$. Then, result (\romannumeral1) can be achieved by induction.

 For (\romannumeral2), we first prove that the tracking error for the nominal system converges to the origin. Then we show that the state of the real system converges to an invariant set along a trajectory lying in the tube $\mathbb{T}$, the center of which is the trajectory of the nominal system. The Lyapunov function for the nominal system is chosen as
 \begin{equation}\label{Lyapunov_tube}
  V(t_k) = J(\tilde{p}_{rf}^{*}(t_k), \tilde{u}_{rf}^{*}(t_k)).
\end{equation}
Consider the difference of the Lypunov function at $t_k$ and $t_{k+1}$,
\begin{eqnarray}\label{Delta V tube}
  \Delta V &=& V(t_{k+1}) - V(t_k) \nonumber\\
   &\leq&  J(\tilde{p}_{rf}(t_{k+1}), \tilde{u}_{rf}(t_{k+1})) - J(\tilde{p}_{rf}^*(t_k), \tilde{u}_{rf}^*(t_k)) \nonumber\\
   &=& \int_{t_{k+1}}^{t_{k+1}+T}(\|\tilde{p}_{rf}(\tau|t_{k+1})\|_Q^2+\|\tilde{u}_{rf} (\tau|t_{k+1})\|_P^2) d\tau \nonumber\\
   & &-\int_{t_{k}}^{t_k+T}(\|\tilde{p}_{rf}^*(\tau|t_{k})\|_Q^2+\|\tilde{u}_{rf}^* (\tau|t_{k})\|_P^2) d\tau \nonumber\\
   & &+ \|\tilde{p}_{rf}(t_{k+1}+T|t_{k+1})\|_R^2- \|\tilde{p}_{rf}^{*}(t_{k}+T|t_{k})\|_R^2 \nonumber\\
   &=& -\int_{t_{k}}^{t_{k+1}}(\|\tilde{p}_{rf}^*(\tau|t_{k})\|_Q^2+\|\tilde{u}_{rf}^* (\tau|t_{k})\|_P^2) d\tau \nonumber\\
   && + \int_{t_{k}+T}^{t_{k+1}+T}(\|\tilde{p}_{rf}(\tau|t_{k+1})\|_Q^2+\|\tilde{u}_{rf} (\tau|t_{k+1})\|_P^2) d\tau \nonumber\\
   &&+ \|\tilde{p}_{rf}(t_{k+1}+T|t_{k+1})\|_R^2- \|\tilde{p}_{rf}^*(t_{k}+T|t_{k})\|_R^2.\nonumber\\
\end{eqnarray}
 By integrating (\ref{stability condition}) form $t_k+T$ to $t_{k+1}+T$, it follows that
\begin{eqnarray}\label{integraing t_k+T t_{k+1}+T}
 \|\tilde{p}_{rf}(t_{k+1}+T|t_{k+1})\|_R^2- \|{p}_f^{e*}(t_{k}+T|t_{k})\|_R^2 \nonumber\\
 + \int_{t_k+T}^{t_{k+1}+T}L(\tilde{p}_f^{e}(\tau), \tilde{u}_{rf}(\tau))d\tau \leq 0.
\end{eqnarray}
Substituting (\ref{integraing t_k+T t_{k+1}+T}) into (\ref{Delta V tube}), we have $\Delta V \leq 0$, which implies that the tracking error for the nominal system converges to the origin asymptotically.

Due to the asymptotic stability of the nominal system, there exists a $\mathcal{KL}$ function $\beta(\cdot, t)$, such that
\begin{equation}\label{}
  \|\tilde{p}_{rf}^*(t)\|\leq\beta(\|\tilde{p}_{rf}^*(t_0)\|, t), \quad \forall t > t_0.
\end{equation}
Furthermore, because of $p_{fe} \in \mathbb{P}_{fe}$ for all $t>t_0$, there exists a $\mathcal{K}$ function $\gamma(\cdot)$ such that
\begin{equation}\label{}
  \|p_{fe}(t)\| \leq \gamma(\eta), \quad \forall t > t_0.
\end{equation}
It follows from $p_{fr}(t) = R(\theta_f)(\tilde{p}_{fr}^*(t)+p_{fe}(t))$ and $p_{fe}(0) = 0$ that
\begin{equation}\label{}
  \|p_{fr}(t)\|\leq\beta(\|{p}_{fr}(t_0)\|, t) + \gamma(\eta).
\end{equation}
Therefore, the solution of system (\ref{error dynamics}) is asymptotically ultimately bounded with Algorithm \ref{algorithm1} and the closed-loop system is ISS.
\epf

\section{NRMPC}\label{sec NRMPC}

In this section, an NRMPC strategy is developed. The state of the nominal system is updated by the actual state at each sampling instant. Unlike tube-MPC, the control sequence obtained is optimal with respect to the current actual state, and only the first control action of the sequence is applied to the real system. The optimization problem of the NRMPC strategy is defined as follows:

\begin{problem}\label{OP2}
\begin{eqnarray}
  \min_{\tilde{u}_f(\tau|t_k)} & & J(\tilde{p}_{rf}(t_k), \tilde{u}_{rf}(t_k)),\\
  s.t. & &\tilde{\xi}_{fh}(t_k|t_k) = \xi_{fh}(t_k), \\
  & &\dot{\tilde{\xi}}_{fh}(\tau|t_k) = f_{h}(\tilde{\xi}_{fh}(\tau|t_k), \tilde{u}_f(\tau|t_k)), \\
   & & \tilde{u}_f(\tau|t_k) \in \mathbb{U},  \\
   & &\| \tilde{p}_{rf}(\tau|t_k)\| \leq \frac{rT}{\tau - t_k},\\
   & & \tilde{p}_{rf}(t_k+T|t_k) \in \Omega_{\varepsilon}, \label{s.t.2}
\end{eqnarray}
where $r = \frac{a(1-\lambda_r)}{\sqrt{\tilde{k}_1^2+\tilde{k}_2^2}}$, $\Omega_{\varepsilon}=\{\tilde{p}_{rf}: \|\tilde{p}_{rf}\| \leq \varepsilon\}$, and $\varepsilon\ < r $.
\end{problem}

Problem \ref{OP2} yields a minimizing control sequence over the interval $[t_k, t_k+T]$ of the same form as in (\ref{optimal control seq}) as well as a minimizing trajectory as in (\ref{optimal trajectory seq}). The control input over $[t_k, t_{k+1}]$ is chosen as
\begin{equation}\label{NRMPC controller}
  u_f(t_k) = \tilde{u}_f^*(t_k|t_k).
\end{equation}

The NRMPC strategy is then described in Algorithm \ref{algorithm2}.
\begin{algorithm}
\caption{NRMPC}\label{algorithm2}
\begin{algorithmic}[1]
\State At time $t_k$, initialize the nominal system state by the actual one $\tilde{\xi}_{fh}(t_k) = \xi_{fh}(t_k)$.
\State Solve Problem \ref{OP2} based on the nominal system to obtain the minimizing control sequence $\bm{\tilde{u}}_f^*(t_k) = \arg\min_{\tilde{u}_f(\tau|t_k)}J(\tilde{p}_{rf}(t_k), \tilde{u}_{rf}(t_k))$.
\State Apply the first portion of the sequence to the real system, i.e., $u_f(t_k)=\tilde{u}_f^*(t_k|t_k))$, during the interval $[t_k, t_{k+1}]$.
\State Update the state of the real system with ${\xi}_{fh}(t_{k+1})$.
\State Update the time instant $t_k = t_{k+1}$ and go to step 1.
\end{algorithmic}
\end{algorithm}

\begin{remark}
For Problem \ref{OP2}, the state of nominal system is updated by the actual one at each step. As a result, the optimization problem has to be solved on-line. However, such an updating strategy yields an optimal control with respect to the current state. The scheme has the same computational burden as the deterministic MPC.
\end{remark}

\begin{remark}
Note that the input domain of NRMPC is larger than that of tube-MPC. NRMPC may therefore have better tracking capability.
\end{remark}

\begin{remark}
As shown in Algorithm \ref{algorithm2}, an open-loop control action is applied to the real system during each sampling interval. However, the existence of disturbances may lead to an error between the actual trajectory and the optimal prediction. This increases the difficulty of analyzing the recursive feasibility using the conventional methods for MPC.
\end{remark}

The following two lemmas guarantee recursive feasibility of Problem \ref{OP2}. Lemma~\ref{lemma feasibility of u} states the existence of the control sequence that is able to drive the state of the nominal system into $\Omega_\varepsilon$ in prediction horizon $T$, and Lemma~\ref{lemma feasibility of trajectory constraint} shows that the state constraint is satisfied by employing that control sequence to the nominal system.

\begin{lemma}\label{lemma feasibility of u}
For the tracking control system (\ref{error dynamics}), assume that there exists an optimal control sequence $\bm{\tilde{u}}_f^*(t_k)$ at instant $t_k$ such that $\tilde{p}_{rf}(t_k+T|t_k)\in \Omega_\varepsilon$, and apply the first control of the sequence to the perturbed system (\ref{disturbed follower system}). Then, there exists a control sequence $\bm{\tilde{u}}_f(t_{k+1})$ at $t_{k+1}$ such that $\tilde{p}_{rf}(t_{k+1}+T|t_{k+1})\in \Omega_\varepsilon$, if
\begin{eqnarray}
  \eta\leq \frac{ e^{-aT}}{\delta}\left(r-\varepsilon\right), \label{bar_w upper bound}\quad
  \tilde{k}\delta \geq \ln\frac{r}{\varepsilon}, \label{sigma upper bound}
\end{eqnarray}
where $\tilde{k} = \min\{\tilde{k}_1, \tilde{k}_2\}$.
\end{lemma}

\bpf
Since Problem \ref{OP2} is feasible at $t_k$, applying the first control of the sequence $\bm{\tilde{u}}_f^*$ during the interval $(t_k, t_{k+1}]$ to the real system may lead to a difference of the trajectory between the actual system and the nominal system. At $t_{k+1}$, this difference is bounded by
\begin{eqnarray}\label{difference at t_{k+1}}
  &&\|\xi_{fh}(t_{k+1}) - \tilde{\xi}_{fh}^*(t_{k+1}|t_k)\| \nonumber\\
  &=& \|\xi_{fh}(t_k) + \int_{t_k}^{t_{k+1}} \left[f_{h}(\xi_{fh}(\tau), \tilde{u}_f^*(\cdot)) + d(\tau)\right] d\tau \nonumber\\
  & & -\tilde{\xi}_{fh}^*(t_k|t_k) - \int_{t_k}^{t_{k+1}} f_{h}(\tilde{\xi}_{fh}^*(\tau|t_k), \tilde{u}_f^*(\cdot)) d\tau\| \nonumber\\
   &\leq&   \int_{t_k}^{t_{k+1}}\left[\|f_{h}(\xi_{fh}(\tau), \tilde{u}_f^*(\cdot))- f_{h}(\tilde{\xi}_{fh}^*(\tau|t_k), \tilde{u}_f^*(\cdot))\|\right]d\tau\nonumber\\
   && +\int_{t_k}^{t_{k+1}}\|d(\tau)\| d\tau \nonumber\\
   &\leq& \eta\delta + a\int_{t_k}^{t_{k+1}} \|\xi_{fh}(\tau) - \tilde{\xi}_{fh}^*(\tau|t_k)\| d\tau \nonumber\\
   &\leq& \eta\delta e^{a\delta}.
\end{eqnarray}
 We construct a feasible control sequence at $t_{k+1}$ for the nominal system as follows.
\begin{equation}\label{feasible input 2}
      \tilde{u}_f(\tau|t_{k+1}) =
      \left\{
      \begin{array}{ll}
       \tilde{u}_f^*(\tau|t_{k}), & \hbox{$\tau \in (t_{k+1}, t_k+T]$,} \\
       \tilde{u}_f^\kappa(\tau|t_k), & \hbox{$\tau \in (t_{k}+T, t_{k+1}+T$]}.
     \end{array}
      \right.
\end{equation}
First, we consider the interval $\tau \in (t_{k+1}, t_k+T]$. Since the state of the nominal system is updated by $\tilde{\xi}(t_{k+1}|t_{k+1}) = \xi(t_{k+1})$, we have
\begin{eqnarray}
&&\|\tilde{\xi}_{fh}(\tau|t_{k+1}) - \tilde{\xi}_{fh}^*(\tau|t_k)\| \nonumber\\
  &=& \|\xi_{fh}(t_{k+1}) + \int_{t_{k+1}}^{\tau} f_{h}(\tilde{\xi}_{fh}(s|t_{k+1}), \tilde{u}_f^*(s|t_k))ds \nonumber\\
  & &- \tilde{\xi}_{fh}^*(t_{k+1}|t_k) - \int_{t_{k+1}}^{\tau} f_{h}(\tilde{\xi}_{fh}^*(s|t_{k}), \tilde{u}_f^*(s|t_k))ds\|.  \nonumber\\
  &\leq& \eta\delta e^{a\delta} + a\int_{t_{k+1}}^{\tau}\|\tilde{\xi}_{fh}(s|t_{k+1}) - \tilde{\xi}_{fh}^*(s|t_k)\| ds.
\end{eqnarray}
Applying Gr{\"o}nwall-Bellman inequality yields
\begin{equation}\label{difference pe}
  \|\tilde{\xi}_{fh}(\tau|t_{k+1}) - \tilde{\xi}_{fh}^*(\tau|t_k)\| \leq \eta\delta e^{a(\tau - t_{k+1}+\delta)}.
\end{equation}
Substituting $t_k+T$ into (\ref{difference pe}) leads to
\begin{equation}\label{}
  \|\tilde{\xi}_{fh}(t_k+T|t_{k+1}) - \tilde{\xi}_{fh}^*(t_k+T|t_k)\| \leq \eta\delta e^{aT}.
\end{equation}
Due to the fact $\|\tilde{p}_{rf}(t_k+T|t_{k+1}) - \tilde{p}_{rf}^*(t_k+T|t_k)\|\leq\|\tilde{\xi}_{fh}(t_k+T|t_{k+1}) - \tilde{\xi}_{fh}^*(t_k+T|t_k)\|$ and the application of triangle inequality, we arrive at
\begin{equation}\label{}
  \|\tilde{p}_{rf}(t_k+T|t_{k+1})\| \leq \|\tilde{p}_{rf}^*(t_k+T|t_k)\| + \eta\delta e^{aT}.
\end{equation}
Since $\tilde{p}_{rf}^*(t_k+T|t_k)\in\Omega_\varepsilon$, i.e. $\|\tilde{p}_{rf}^*(t_k+T|t_k)\| \leq \varepsilon$, and $\eta\leq \frac{ e^{-aT}}{\delta}(r-\varepsilon)$, we obtain
\begin{equation}\label{terminal feasible state in Omega_f}
  \|\tilde{p}_{rf}(t_k+T|t_{k+1})\| \leq r,
\end{equation}
which implies $\tilde{p}_{rf}(t_k+T|t_{k+1})\in \Omega$.

Next, consider the interval $\tau \in (t_k+T, t_{k+1}+T]$, during which the local controller $\tilde{u}_f^\kappa(\tau|t_k)$ is applied to the nominal system
\begin{eqnarray*}
  \frac{d}{d\tau}\|\tilde{p}_{rf}(\tau|t_{k+1})\|^2 &=& -2(\tilde{k}_1{\tilde{x}_{rf}}^2+\tilde{k}_2{\tilde{y}_{rf}}^2)\\
  &\leq& -2\tilde{k}\|\tilde{p}_{rf}(\tau|t_{k+1})\|^2.
\end{eqnarray*}
Applying the comparison principle yields
\begin{eqnarray*}
\|\tilde{p}_{rf}(t_{k+1}+T|t_{k+1})\| \leq \|\tilde{p}_{rf}(t_k+T|t_{k+1})\|e^{-\delta\tilde{k}}.
\end{eqnarray*}
It follows from $\tilde{k}\delta \geq \ln \displaystyle\frac{r}{\varepsilon}$ that
\begin{equation}\label{}
  \|\tilde{p}_{rf}(t_{k+1}+T|t_{k+1})\| \leq \varepsilon.
\end{equation}
This proves the existence of a control sequence at $t_{k+1}$ which is able to drive the tracking error of the nominal system into the terminal region $\Omega_\varepsilon$.
\epf

\begin{lemma}\label{lemma feasibility of trajectory constraint}
For the tracking control system (\ref{error dynamics}), assume that there exists an optimal control sequence $\bm{\tilde{u}}_f^*(t_k)$ at instant $t_k$ such that the trajectory constraint is satisfied, i.e., $\tilde{p}_{fr}^*(\tau|t_k)\leq\frac{rT}{\tau - t_k}$, and apply the first control of the sequence to the perturbed system (\ref{disturbed follower system}). Then, at $t_{k+1}$, by the control sequence (\ref{feasible input 2}), the trajectory constraint is also satisfied, if the parameter $\varepsilon$ satisfies
\begin{equation}\label{varepsilon condition}
\varepsilon\geq \frac{r(T-\delta)}{T}.
\end{equation}
\end{lemma}

\bpf
To prove $\|\tilde{p}_{fr}(\tau|t_{k+1})\|\leq\frac{rT}{\tau - t_{k+1}}$, $\tau \in (t_{k+1}, t_{k+1} + T]$, we first consider the interval $\tau \in (t_{k+1}, t_k + T]$. From (\ref{difference pe}), it follows that
\begin{equation}\label{}
  \|\tilde{p}_{fr}(\tau|t_{k+1})\| \leq \|\tilde{p}_{rf}^*(\tau|t_k)\|+ \eta\delta e^{aT}.
\end{equation}
Due to (\ref{bar_w upper bound}) and $\tilde{p}_{fr}^*(\tau|t_k)\leq\frac{T}{\tau - t_k}$, we obtain
\begin{equation}\label{pe constraint}
  \|\tilde{p}_{fr}(\tau|t_{k+1})\| \leq \frac{rT}{\tau - t_k}+ (r-\varepsilon).
\end{equation}
From (\ref{varepsilon condition}), we have
\begin{equation}\label{epsilon1}
 r - \varepsilon \leq \frac{\delta r}{T-\delta}\leq\frac{\delta rT}{(\tau-t_{k+1})(\tau-t_k)}.
\end{equation}
Substituting (\ref{epsilon1}) into (\ref{pe constraint}), we obtain
\begin{equation}
  \|\tilde{p}_{fr}(\tau|t_{k+1})\|\leq\frac{rT}{\tau - t_{k+1}},
\end{equation}
which proves that the state constraint is satisfied over the interval $\tau \in [t_{k+1}, t_{k}+T]$.

Next, consider the interval $\tau \in [t_k+T, t_{k+1}+T]$. By Lemma~\ref{lemma feasibility of u}, it holds that $\|\tilde{p}_{rf}(t_k+T|t_{k+1})\| \leq r$ once $\|\tilde{p}_{rf}^*(t_k+T|t_k)\| \leq \varepsilon$. Since $\frac{rT}{\tau - t_{k+1}} \geq r$, $\|\tilde{p}_{fr}(\tau|t_{k+1})\|\leq\frac{rT}{\tau - t_{k+1}}$ is naturally satisfied over the interval $\tau \in [t_k+T, t_{k+1}+T]$, thereby completing the proof.
\epf

\begin{thm}\label{theorem NRMPC}
For the tracking control system~(\ref{error dynamics}), suppose that Problem~\ref{OP1} is feasible at time $t_0$ and the parameters satisfy the conditions in Lemma~ \ref{lemma feasibility of u} and Lemma~\ref{lemma feasibility of trajectory constraint}. Then,
\begin{enumerate}
  \item[(\romannumeral1)] Problem~\ref{OP1} is feasible for all $t > t_0$;
  \item[(\romannumeral2)] the tracking control system (\ref{error dynamics}) is ISS if
  \begin{eqnarray}\label{stability parameters condition}
     \underline{q}\varepsilon^2>  \frac{1}{2}\eta e^{aT}(r+\varepsilon)+ \frac{q^2\eta^2\delta}{2a}(e^{2aT} - e^{2a\delta}) \nonumber\\
    + \frac{2q^2\eta r}{\sqrt{2}a}(\frac{T^2}{\delta}-T)^{\frac{1}{2}}(e^{2aT} - e^{2a\delta})^{\frac{1}{2}},
  \end{eqnarray}
  where $\underline{q} = \min\{q_1, q_2\}$.
\end{enumerate}
\end{thm}

\bpf
(\romannumeral1) Assume Problem \ref{OP2} is feasible at instant $t_k$, then feasibility of Problem \ref{OP2} at $t_{k+1}$ implies the existence of a control sequence that is able to drive the state of the nominal system to the terminal region $\Omega_\varepsilon$ while satisfying all the constraints. In terms of Algorithm \ref{algorithm2}, the first control of the optimal control sequence is applied to the system. From Lemma~\ref{lemma feasibility of u}, at $t_{k+1}$, a feasible control sequence in (\ref{feasible input 2}) renders $\tilde{p}_{rf}(t_{k+1}+T|t_{k+1})\in \Omega_\varepsilon$ while satisfying $\tilde{u}_f(\tau|t_{k+1})\in \Omega$ for $\tau \in [t_{k+1}, t_{k+1}+T]$. Meanwhile, From Lemma~\ref{lemma feasibility of trajectory constraint}, by applying the control sequence (\ref{feasible input 2}) to the nominal system (\ref{nominal follower system}), the trajectory constraint is satisfied, i.e., $\| \tilde{p}_{rf}(\tau|t_{k+1})\| \leq \frac{rT}{\tau - t_{k+1}}$, implying the feasibility of Problem \ref{OP2} at $t_{k+1}$. Hence, the feasibility of Problem \ref{OP2} at the initial time $t_0$ results in the feasibility for all $t > t_0$ by induction.

(\romannumeral2) Choose a Lyapunov function as follows
\begin{equation}\label{Lyapunov}
  V(p_{rf}(t_k)) = J(\tilde{p}_{rf}^*(t_k), \tilde{u}_{rf}^*(t_k)).
\end{equation}

According to Riemann integral principle, there exists a constants $0<c_1\leq\delta$ such that
\begin{equation}\label{}
  V(p_{rf}) \geq c_1L(t_k,, u_{rf})\triangleq \alpha_1(\|p_{rf}\|),
\end{equation}
where $\alpha_1(\cdot)$ is obviously a $\mathcal{K}_\infty$ function.

On the other hand, from (\ref{stability condition}), we have
$$V(p_{rf}(t_k)) \leq g(p_{rf}(t_k))+g(p_{rf}(t_k+T|t_k)), \forall p_{rf}\in\Omega_\varepsilon.$$
Due to the decreasing property of $g(\cdot)$ in $\Omega_\varepsilon$ with respect to time, it follows that
\begin{equation}\label{}
  V(p_{rf}(t_k)) \leq 2g(p_{rf}(t_k)),\quad \forall p_{rf}\in\Omega_\varepsilon.
\end{equation}

Because the origin lies in the interior of $\Omega_\varepsilon$ and $2g(p_{rf}(t))\leq \varepsilon$, $\forall p_{rf}\in\Omega_\varepsilon$, it holds that $2g(p_{rf}(t))\geq \varepsilon$ if $p_{rf} \in \mathbb{R}^{2\times2}\setminus\Omega_\varepsilon$. Due to the feasibility of Problem \ref{OP2}, there exists an upper-bound $c_2 > \varepsilon$ for $V(p_{rf}(t))$. Thus $\alpha_2(\|p_{rf}\|)= \displaystyle\frac{c_2}{\varepsilon}g(p_{rf}(t))$ is a $\mathcal{K}_\infty$ function such that $\alpha_2(\|p_{rf}\|) \geq c_2$ thereby satisfying $\alpha_2(\|p_{rf}(t)\|) \geq V(p_{rf}(t))$.

This proves the existence of $\mathcal{K}_\infty$ functions $\alpha_1(\cdot)$ and $\alpha_2(\cdot)$ satisfying
\begin{equation}\label{}
  \alpha_1(\|p_{rf}(t)\|)\leq V(p_{rf}(t))\leq\alpha_2(\|p_{rf}(t)\|).
\end{equation}

The difference of the value Lypunov function at $t_k$ and $t_{k+1}$ satisfies
\begin{eqnarray}\label{Delta V}
  \Delta V &=& V(p_{rf}(t_{k+1})) - V(p_{rf}(t_k)) \nonumber\\
   &\leq& J(\tilde{p}_{rf}(t_{k+1}), \tilde{u}_{rf}(t_{k+1})) - J(\tilde{p}_{rf}^*(t_k), \tilde{u}_{rf}^*(t_k))\nonumber\\
   &\triangleq& \Delta V_1 + \Delta V_2 + \Delta V_3,
\end{eqnarray}
in which
\begin{eqnarray*}
  \Delta V_1 &=& \int_{t_{k+1}}^{t_k+T}(\|\tilde{p}_{rf}(\tau|t_{k+1})\|_Q^2 - \|\tilde{p}_{rf}^*(\tau|t_{k})\|_Q^2) d\tau, \\
  \Delta V_2 &=& \int_{t_{k}+T}^{t_{k+1}+T}(\|\tilde{p}_{rf}(\tau|t_{k+1})\|_Q^2+\|\tilde{u}_{rf} (\tau|t_{k+1})\|_P^2) d\tau \\
  &&+ \|\tilde{p}_{rf}(t_{k+1}+T|t_{k+1})\|_R^2 - \|\tilde{p}_{rf}^*(t_{k}+T|t_{k})\|_R^2,\\
  \Delta V_3 &=& -\int_{t_{k}}^{t_{k+1}}(\|\tilde{p}_{rf}^*(\tau|t_{k})\|_Q^2+\|\tilde{u}_{rf}^* (\tau|t_{k})\|_P^2) d\tau.
\end{eqnarray*}

For $\Delta V_1$, it holds that
\begin{eqnarray}\label{}
\Delta V_1 &\leq& \int_{t_{k+1}}^{t_k+T} (\|\tilde{p}_{rf}(\tau|t_{k+1})-\tilde{p}_{rf}^*(\tau|t_{k})\|_Q)\nonumber\\
&&\times(\|\tilde{p}_{rf}(\tau|t_{k+1})\|_Q
  + \|\tilde{p}_{rf}^*(\tau|t_{k})\|_Q) d\tau.
\end{eqnarray}
By (\ref{difference pe}), we have
\begin{eqnarray}
\Delta V_1&\leq& \int_{t_{k+1}}^{t_k+T} [q^2\eta\delta e^{a(\tau + \delta - t_{k+1})}(2\|\tilde{p}_{rf}^*(\tau|t_{k})\| \nonumber\\
    &&  +\eta\delta e^{a(\tau + \delta - t_{k+1})})]d\tau \nonumber\\
&=& \int_{t_{k+1}}^{t_k+T} 2 q^2\eta\delta e^{a(\tau + \delta - t_{k+1})} \|\tilde{p}_{rf}^*(\tau|t_{k})\| \nonumber\\
  &&+ q^2\eta^2\delta^2 e^{2a(\tau + \delta - t_{k+1})}d\tau \nonumber\\
  &\leq&\int_{t_{k+1}}^{t_k+T} 2 q^2\eta\delta e^{a(\tau + \delta - t_{k+1})} \|\tilde{p}_{rf}^*(\tau|t_{k})\| d\tau \nonumber\\
  &&+ \frac{q^2\eta^2\delta^2}{2a}(e^{2aT} - e^{2a\delta}).
\end{eqnarray}
Applying H\"{o}lder inequality to the first term of the last inequality yields
\begin{eqnarray}\label{Delta V1b}
  \Delta V_1&\leq&\left(\int_{t_{k+1}}^{t_k+T}\|\tilde{p}_{rf}^*(\tau|t_{k})\|^2 d\tau\right)^{\frac{1}{2}} \frac{2q^2\eta\delta}{\sqrt{2}a}(e^{2aT} - e^{2a\delta})^{\frac{1}{2}} \nonumber\\
    &&+ \frac{q^2\eta^2\delta^2}{2a}(e^{2aT} - e^{2a\delta})\nonumber\\
    &\leq& \frac{2q^2\eta\delta r}{\sqrt{2}a}(\frac{T^2}{\delta}-T)^{\frac{1}{2}}(e^{2aT} - e^{2a\delta})^{\frac{1}{2}} \nonumber\\
    &&+ \frac{q^2\eta^2\delta^2}{2a}(e^{2aT} - e^{2a\delta}).
\end{eqnarray}

Rewrite $\Delta V_2$ as
\begin{eqnarray}\label{Delta V2a}
  \Delta V_2 &=&\!\!\! \int_{t_k+T}^{t_{k+1}+T} (\|\tilde{p}_{rf}(\tau|t_{k+1})\|_Q^2 + \|\tilde{u}_{rf}(\tau|t_{k+1})\|_P^2 d)\tau\nonumber\\
   &&\!\!+ \|\tilde{p}_{rf}(t_{k+1}+T|t_{k+1})\|_R^2 - \|\tilde{p}_{rf}^*(t_{k}+T|t_{k})\|_R^2\nonumber\\
  &&\!\!\!\!\! +  \|\tilde{p}_{rf}(t_{k}+T|t_{k+1})\|_R^2 - \|\tilde{p}_{rf}(t_{k}+T|t_{k+1})\|_R^2.
\end{eqnarray}
Integrating (\ref{stability condition}) from $t_k+T$ to $t_{k+1}+T$ and substituting it into (\ref{Delta V2a}) leads to
\begin{eqnarray}\label{Delta V2b}
  \Delta V_2 &\leq& \|\tilde{p}_{rf}(t_{k}+T|t_{k+1})\|_R^2 - \|\tilde{p}_{rf}^*(t_{k}+T|t_{k})\|_R^2 \nonumber\\
   &\leq& (\frac{1}{2}\|\tilde{p}_{rf}(t_{k}+T|t_{k+1})-\tilde{p}_{rf}^*(t_{k}+T|t_{k})\|)\nonumber\\
   &&\times\left(\|\tilde{p}_{rf}(t_{k}+T|t_{k+1})\|+\|\tilde{p}_{rf}^*(t_{k}+T|t_{k})\|\right)\nonumber\\
   &\leq& \frac{1}{2}\eta\sigma e^{aT}(\varepsilon + r).
\end{eqnarray}

For $\Delta V_3$, we first assume $\|\tilde{p}_{rf}(t_{k+1}|t_k)\|> \varepsilon$, which implies $\|\tilde{p}_{rf}(\tau|t_k)\|> \varepsilon$, $\tau \in (t_k, t_{k+1}]$, and thus we obtain
 \begin{eqnarray}\label{Delta V3b}
    \Delta V_3 &<& -\int_{t_{k}}^{t_{k+1}}\|\tilde{p}_{rf}^*(\tau|t_{k})\|_Q^2d\tau
          \leq -\underline{q}\delta\varepsilon^2.
  \end{eqnarray}

In combination with (\ref{Delta V1b}),(\ref{Delta V2b}) and (\ref{Delta V3b}), the inequality (\ref{Delta V}) thus satisfies
\begin{eqnarray}
    \Delta V &<& -\underline{q}\delta\varepsilon^2 + \frac{1}{2}\eta\delta e^{aT}(r+\varepsilon)+ \frac{q^2\eta^2\delta^2}{2a}(e^{2aT} - e^{2a\delta}) \nonumber\\
    &&+ \frac{2q^2\eta\delta r}{\sqrt{2}a}(\frac{T^2}{\delta}-T)^{\frac{1}{2}}(e^{2aT} - e^{2a\delta})^{\frac{1}{2}}.
\end{eqnarray}
From (\ref{stability parameters condition}), $\Delta V<0$ holds. It follows from Theorem 2 of \cite{michalska1993robust} that $\|\tilde{p}_{rf}^*(t_k|t_k)\|\leq \varepsilon$ for $t_k\geq t_f$, where $t_f>t_0$ is a finite time instant. When the tracking error enters into the terminal region, i.e., $p_{rf}(t_k) \in \Omega_\varepsilon$, reconsider $\Delta V_1$ and $\Delta V_3$:
 \begin{eqnarray}\label{Delta V1n}
   \Delta V_1  &\leq& \int_{t_{k+1}}^{t_k+T} 2 q^2\eta\delta \varepsilon e^{a(\tau - t_{k+2})} d\tau
  + \frac{q^2\eta^2\delta^2}{2a}(e^{2aT} - e^{2a\delta})\nonumber\\
  &=&\frac{2 q^2\eta\delta\varepsilon}{a}(e^{aT} - e^{a\delta})+ \frac{q^2\eta^2\delta^2}{2a}(e^{2aT} - e^{2a\delta}).\nonumber
 \end{eqnarray}

Due to the decreasing property of $\|\tilde{p}_{rf}^*(\tau|t_{k})\|_Q^2$ in $\Omega_\varepsilon$, it follows that
\begin{eqnarray*}
    \Delta V_3 &\leq& -\underline{q}\delta\|{p}_{rf}^*(t_{k+1}|t_{k})\|^2.
  \end{eqnarray*}
Since $\|\tilde{p}_{rf}^*(\tau|t_{k+1})\| \leq \|\tilde{p}_{rf}^*(\tau|t_k)\| + \eta\delta e^{aT}$, we have $\|{p}_{rf}(t_{k+1})\|^2 \leq \|p_{rf}^*(t_{k+1}|t_k)\|^2+ \eta^2\delta^2 e^{2a\delta} + 2\varepsilon\eta\delta e^{a\delta}$. Consequently,
\begin{eqnarray*}
\Delta V_3 \leq -\underline{q}\delta\|{p}_{rf}(t_{k+1})\|^2 + \underline{q}\eta^2\delta^3 e^{2a\delta} + 2\underline{q}\varepsilon\eta\delta^2 e^{a\delta}.
\end{eqnarray*}

As a result, it holds that
\begin{equation}\label{}
\Delta V \leq -\underline{q}\delta\|{p}_{rf}(t_{k+1})\|^2 + \sigma(\eta),
\end{equation}
where $\sigma(\eta) = \frac{2 q^2\eta\delta\varepsilon}{a}(e^{aT} - e^{a\delta})+ \frac{q^2\eta^2\delta^2}{2a}(e^{2aT} - e^{2a\delta}) + \frac{1}{2}\eta\sigma e^{aT}(\varepsilon + r) + \underline{q}\eta^2\delta^3 e^{2a\delta} + 2\underline{q}\varepsilon\eta\delta^2 e^{a\delta}$  is obviously a $\mathcal{K}$ function with respect to $\eta$. Hence, the theorem is proved.
\epf

\section{Simulation results}\label{sec simulation}
The simulation is implemented on a PC equipped with a dual-core 3.20 GHz Intel i5 CPU, 7.88 GB RAM and 64-bit Windows 10 operating system. The optimization problem is transcribed by Tool Box ICLOCS (Imperical College London Optimal Control Software, see \cite{ICLOCS}), 1.2 version, and solved by NLP (Nonlinear Programming) solver IPOPT (Interior Point OPTimizer, see \cite{wachter2006implementation}), 3.11.8 version.

The mechanism parameters of the two homogeneous robots used in the simulation are taken from an educational robot named E-puck \cite{mondada2009puck}, and are given by $a = 0.13$ m/s, $\rho = 0.0267$~m and $b = a/\rho = 4.8598$~rad/s. The trajectory to be tracked is a circular motion with linear velocity $v_r = 0.015$~m/s, angular velocity $\omega_r = 0.04$~m/s and initial configuration $\xi_r(0) = [0, 0, \frac{\pi}{3}]^\mathrm{T}$. The initial configuration of the follower is set to be $\xi_{fh} = [0.2, -0.2, -\frac{\pi}{2}]^\mathrm{T}$. The disturbances are bounded by $\eta = 0.004$. For the tracking objective, the prediction horizon and the sampling period are set to be $T = 2$ s and $\delta = 0.2$~s, respectively. The positive define matrices $P$ and $Q$ are chosen, according to Lemma~\ref{lemma terminal region}, as $P = \mathrm{diag}\{0.4, 0.4\}$ and $Q = \mathrm{diag}\{0.2, 0.2\}$, respectively. The feedback gains for the terminal controller are set to be $\tilde{k}_1 = \tilde{k}_2 = 1.2$ to satisfy the requirements given by Lemma~\ref{lemma terminal region}.

\begin{figure}[htb]
  \centering
  \includegraphics[width=9cm]{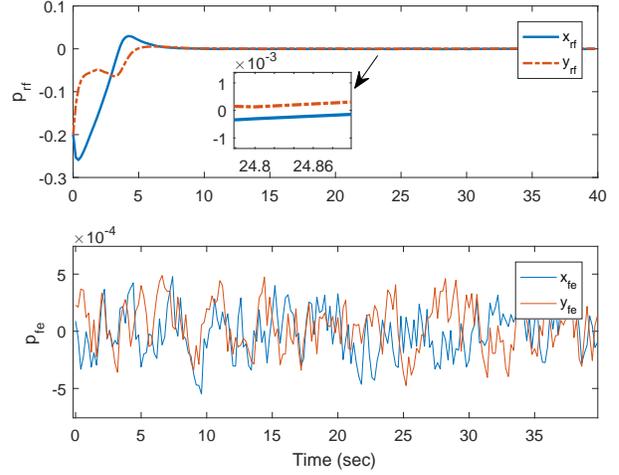}\\
  \caption{Tracking errors by using tube-MPC.}\label{SIM_Tube_MPC_tracking_errors}
\end{figure}

\begin{figure}[htb]
  \centering
  \includegraphics[width=9cm]{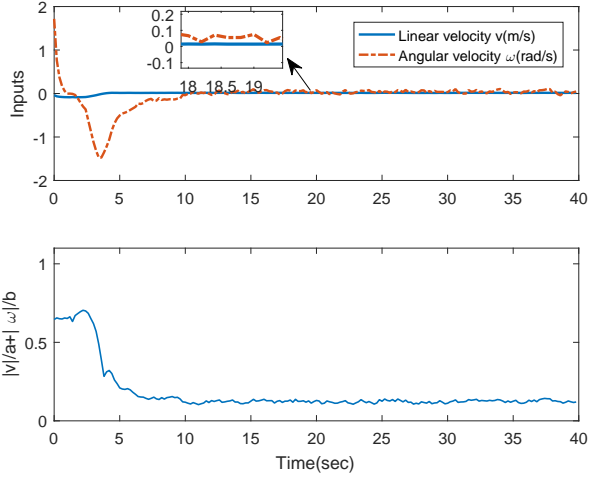}\\
  \caption{Control input of tube-MPC.}\label{SIM_Tube_MPC_input}
\end{figure}

First, let us design tube-MPC according to Lemma~\ref{lemma tube} and Theorem~\ref{theorem tube}. We set the feedback gain to be $K = \rm{diag}\{-2.3, -2.3\}$. The control input constraint for Problem \ref{OP1} is $\mathbb{U}_{tube}= \lambda_{tube}\mathbb{U}$ with $\lambda_{tube} = 0.6636$, and the terminal region is given by $\Omega_{tube} = \{\tilde{p}_{rf}: |\tilde{x}_{rf}| + |\tilde{y}_{rf}|\leq 0.0542\}$. To evaluate the tracking performance, we take the tracking error $p_{rf}$ and the real position deviation from the center of the tube $p_{fe}$ as indexes. Applying Algorithm \ref{algorithm1} to the tracking system yields the tracking performance as shown in Fig.~\ref{SIM_Tube_MPC_tracking_errors}. It can be found, from Fig.~\ref{SIM_Tube_MPC_tracking_errors}, that the tracking error converges to a neighbourhood of the origin, and the trajectory of the follower lies in the tube $\mathbb{T}= p_{fh}^* \oplus \mathbb{P}_{fe}$ with $\mathbb{P}_{fe} = \{|p_{fe}|\leq\left[
            \begin{array}{c}
              0.0017 \\
             0.0017  \\ \end{array}
          \right]\}$, which is obtained from Lemma~\ref{lemma tube}.
Fig.~\ref{SIM_Tube_MPC_input} shows the control input performance of the follower. We also take $|v|/a+|\omega|/b$ as an index to evaluate the input constraint. The fluctuated control signal indicates the effectiveness of the feedback part of the controller which reduces the tracking error caused by the disturbances. Furthermore, the input constraint $\mathbb{U}_{tube}$ for the nominal system is active over the interval $t \in [0, 3]$, while the constraint $\mathbb{U}$ for the real system is not active, which indicates the weak control ability.

\begin{figure}[htb]
  \centering
  \includegraphics[width=9cm]{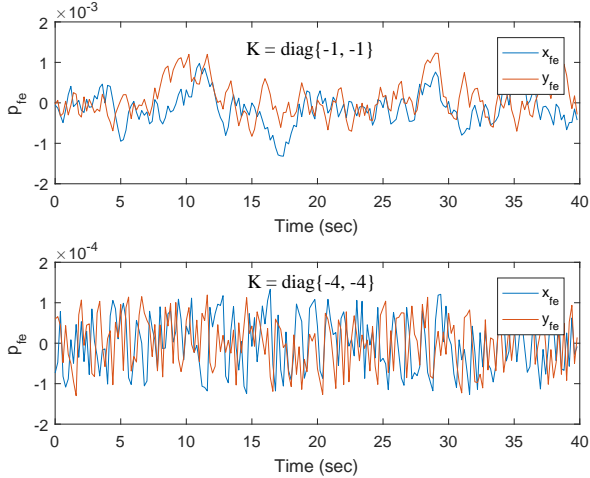}\\
  \caption{Real-time position deviation from the center of the tube $p_{fe}$ with different feedback gains.}\label{SIM_tuning_K}
\end{figure}

 To show the effect of different choices of the feedback gain $K$ on the tracking performance, we set $K = \rm{diag}\{-1, -1\}$ and $K = \rm{diag}\{-4, -4\}$, respectively, to observe the difference between the actual trajectory and the optimal one. As shown in Fig.~\ref{SIM_tuning_K}, increasing of $K$ reduces the difference of the actual position and the center of the tube and therefore improves the tracking performance.

\begin{figure}[htb]
  \centering
  \includegraphics[width=9cm]{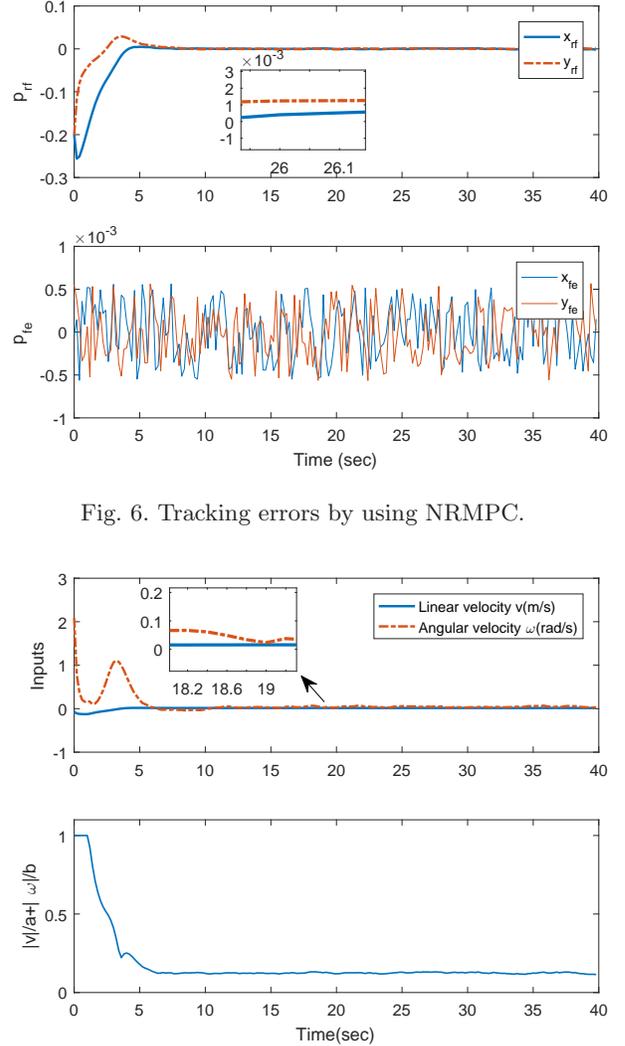}\\
  \caption{Tracking errors by using NRMPC.}\label{SIM_SO_MPC_tracking_errors}
\end{figure}

\begin{figure}[htb]
  \centering
  \includegraphics[width=9cm]{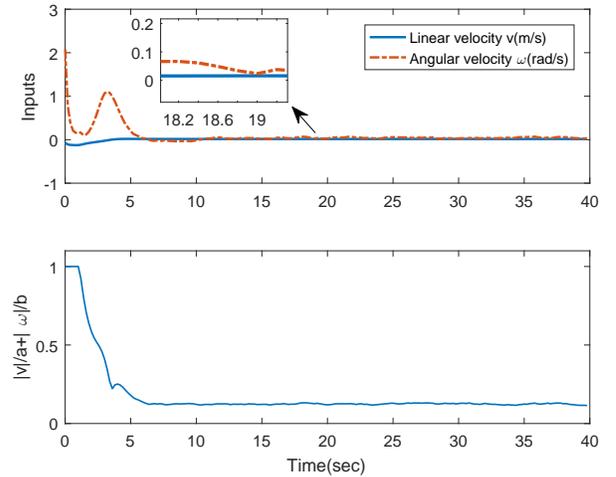}\\
  \caption{Control input of NRMPC.}\label{SIM_SO_MPC_input}
\end{figure}

Next, design NRMPC according to Lemma~\ref{lemma feasibility of u}, Lemma~\ref{lemma feasibility of trajectory constraint} and Theorem~\ref{theorem NRMPC}. The input constraint of NRMPC differs from the constraint of tube-MPC and is given by $\tilde{u}_f \in \mathbb{U}$ according to Algorithm~\ref{algorithm2}. The terminal region is designed as $\Omega_{\varepsilon}=\{\tilde{p}_{rf}: \|\tilde{p}_{rf}\| \leq 0.063\}$, and consequently $\varepsilon = 0.063$, which satisfies the conditions in Lemma~\ref{lemma feasibility of u}, Lemma~\ref{lemma feasibility of trajectory constraint} and Theorem~\ref{theorem NRMPC}. Fig.~\ref{SIM_SO_MPC_tracking_errors} presents the tracking performance of Algorithm~\ref{algorithm2}. It can be observed that the tracking error converges to a neighbourhood of the origin. To compare the influence level by disturbances of the two strategies proposed, we define $p_{fe}(t_k) = p_f(t_k) - \tilde{p}_f^*(t_k)$ in NRMPC. It can be seen that the tracking performance is directly influenced by disturbances due to the open-loop control during each sampling period. Fig.~\ref{SIM_SO_MPC_input} shows the control input under NRMPC. According to Algorithm~\ref{algorithm2}, the control signal at each time instant is optimal corresponding to its current state, which indicates its robustness. We also note that the input constraint is active over the interval $t\in[0, 1.5]$, which demonstrates a better tracking capability.

\begin{figure}[htb]
  \centering
  \includegraphics[width=9cm]{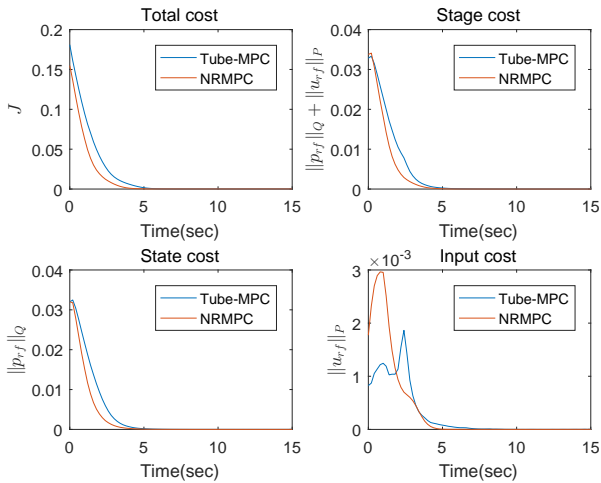}\\
  \caption{Costs of tube-MPC and NRMPC.}\label{SIM_Cost}
\end{figure}

\begin{figure}[htb]
  \centering
  \includegraphics[width=9cm]{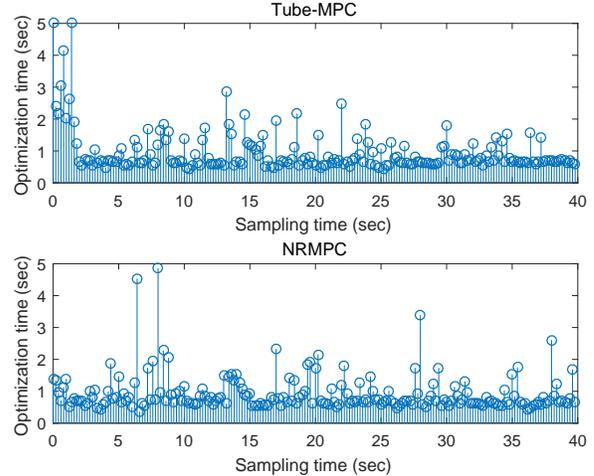}\\
  \caption{Computation time of tube-MPC and NRMPC.}\label{SIM_Optimization_time}
\end{figure}

To further compare tube-MPC with NRMPC, we take cost function $J$, real stage cost $\|p_{rf}\|_Q^2 + \|u_{rf}\|_P^2$, real state cost $\|p_{rf}\|_Q^2$ and real input cost $\|u_{rf}\|_P^2$ to evaluate the converging performance. Their cost curves are plotted in Fig \ref{SIM_Cost}. As it can be seen, the total cost, the stage cost and the state cost decrease faster by implementing NRMPC than by tube-MPC. However, the input cost of NRMPC is higher than that of tube-MPC. This is explained by the fact that the input constraint of tube-MPC is tighter than that of NRMPC, which may degrade the control capability. This also helps explaining why the tracking error decreases faster by NRMPC than by tube-MPC. Fig.~\ref{SIM_Optimization_time} provides the computation time in solving the optimization problems. It shows that there is no significant difference between NRMPC and tube-MPC, which implies that they have almost the same computational complexity. However, as stated in Remark \ref{remark tube off-line}, the optimization problem can be solved off-line in tube-MPC, whereas the optimization problem has to be solved on-line in NRMPC.

Finally, we summarize the simulation study as follows:
\begin{itemize}
  \item [(\romannumeral1)] Tube-MPC presents a better steady state performance than NRMPC. This is because the control strategy of tube-MPC consists of two parts: optimal control and feedback part, while NRMPC is open-loop control during each sampling period.
  \item [(\romannumeral2)] NRMPC performs better in terms of dynamic property than tube-MPC due to the tighter input constraint of tube-MPC.
  \item [(\romannumeral3)] The computational complexities of tube-MPC and NRMPC are almost the same. However, the optimization problem in tube-MPC can be solved off-line, which may enhance its real-time performance.
\end{itemize}

\section{Conclusion}\label{sec conclusion}
In this paper, two robust MPC strategies have been developed for tracking of nonholonomic systems with coupled input constraint and bounded disturbances. We first developed a tube-MPC strategy, where the trajectory of the real system is constrained in a tube centered along the optimal nominal trajectory by a nonlinear feedback law based on the deviation of the actual states from the optimal nominal states. Tube-MPC possesses robustness but sacrifices optimality, thus we further developed the NRMPC scheme, where the state of the nominal system is updated by the actual state at each step. It was shown that the tracking control system is ISS under both robust MPC strategies, and their optimization problems are feasible. Simulation results illustrated the effectiveness of the two schemes and their respective advantages.

%
%

\bibliographystyle{unsrt}
\bibliography{bibfile}

\begin{thebibliography}{10}

\bibitem{simanek2015evaluation}
J.~Simanek, M.~Reinstein, and V.~Kubelka.
\newblock Evaluation of the {EKF-based} estimation architectures for data
  fusion in mobile robots.
\newblock {\em IEEE/ASME Transactions on Mechatronics}, 20(2):985--990, 2015.

\bibitem{wang2014distributed}
P.~Wang and B.~Ding.
\newblock Distributed {RHC} for tracking and formation of nonholonomic
  multi-vehicle systems.
\newblock {\em IEEE Transactions on Automatic Control}, 59(6):1439--1453, 2014.

\bibitem{lafferriere2005decentralized}
G.~Lafferriere, A.~Williams, J.~Caughman, and J.J.P. Veerman.
\newblock Decentralized control of vehicle formations.
\newblock {\em Systems $\&$ control letters}, 54(9):899--910, 2005.

\bibitem{jiangdagger1997tracking}
Z.~P. Jiang and H.~Nijmeijer.
\newblock Tracking control of mobile robots: a case study in backstepping.
\newblock {\em Automatica}, 33(7):1393--1399, 1997.

\bibitem{yang1999sliding}
J.~Yang and J.~Kim.
\newblock Sliding mode control for trajectory tracking of nonholonomic wheeled
  mobile robots.
\newblock {\em IEEE Transactions on Robotics and Automation}, 15(3):578--587,
  1999.

\bibitem{lee2001tracking}
T.~Lee, K.~Song, C.~Lee, and C.~Teng.
\newblock Tracking control of unicycle-modeled mobile robots using a saturation
  feedback controller.
\newblock {\em IEEE Transactions on Control Systems Technology}, 9(2):305--318,
  2001.

\bibitem{marshall2006pursuit}
J.~A. Marshall, M.~E. Broucke, and B.~A. Francis.
\newblock Pursuit formations of unicycles.
\newblock {\em Automatica}, 42(1):3--12, 2006.

\bibitem{ghommam2010formation}
J.~Ghommam, H.~Mehrjerdi, M.~Saad, and F.~Mnif.
\newblock Formation path following control of unicycle-type mobile robots.
\newblock {\em Robotics and Autonomous Systems}, 58(5):727--736, 2010.

\bibitem{gu2006receding}
D.~Gu and H.~Hu.
\newblock Receding horizon tracking control of wheeled mobile robots.
\newblock {\em IEEE Transactions on Control Systems Technology},
  14(4):743--749, 2006.

\bibitem{mayne2000constrained}
D.~Q. Mayne, J.~B. Rawlings, C.~V. Rao, and P.~O.~M. Scokaert.
\newblock Constrained model predictive control: Stability and optimality.
\newblock {\em Automatica}, 36(6):789--814, 2000.

\bibitem{chen2009leader}
J.~Chen, D.~Sun, J.~Yang, and H.~Chen.
\newblock Leader-follower formation control of multiple nonholonomic mobile
  robots incorporating receding-horizon scheme.
\newblock {\em The International Journal of Robotics Research}, 29:727--747,
  2010.

\bibitem{sun2016receding}
Z.~Sun and Y.~Xia.
\newblock Receding horizon tracking control of unicycle-type robots based on
  virtual structure.
\newblock {\em International Journal of Robust and Nonlinear Control}, 2016,
  DOI: 10.1002/rnc.3555.

\bibitem{rawlings2009model}
J.~B. Rawlings and D.~Q. Mayne.
\newblock {\em Model predictive control: Theory and design}.
\newblock Nob Hill Pub., 2009.

\bibitem{scokaert1995stability}
P.~O.~M. Scokaert and J.~B. Rawlings.
\newblock Stability of model predictive control under perturbations.
\newblock In {\em Proceedings of the IFAC Symposium on Nonlinear Control
  Systems Design}, pages 1317--1322, 1995.

\bibitem{marruedo2002stability}
D.~L. Marruedo, T.~Alamo, and E.~F. Camacho.
\newblock Stability analysis of systems with bounded additive uncertainties
  based on invariant sets: Stability and feasibility of {MPC}.
\newblock In {\em Proceedings of American Control Conference}, pages 364--369,
  2002.

\bibitem{kothare1996robust}
M.~V. Kothare, V.~Balakrishnan, and M.~Morari.
\newblock Robust constrained model predictive control using linear matrix
  inequalities.
\newblock {\em Automatica}, 32(10):1361--1379, 1996.

\bibitem{lee1997worst}
J.~H. Lee and Z.~Yu.
\newblock Worst-case formulations of model predictive control for systems with
  bounded parameters.
\newblock {\em Automatica}, 33(5):763--781, 1997.

\bibitem{wan2002robust}
Z.~Wan and M.~V. Kothare.
\newblock Robust output feedback model predictive control using off-line linear
  matrix inequalities.
\newblock {\em Journal of Process Control}, 12(7):763--774, 2002.

\bibitem{magni2003robust}
L.~Magni, G.~De~Nicolao, R.~Scattolini, and F.~Allg{\"o}wer.
\newblock Robust model predictive control for nonlinear discrete-time systems.
\newblock {\em International Journal of Robust and Nonlinear Control},
  13(3-4):229--246, 2003.

\bibitem{chen1997game}
H.~Chen, C.~W. Scherer, and F.~Allg{\"o}wer.
\newblock A game theoretic approach to nonlinear robust receding horizon
  control of constrained systems.
\newblock In {\em Proceedings of American control conference}, volume~5, pages
  3073--3077, 1997.

\bibitem{limon2006input}
D.~Lim{\'o}n, T.~Alamo, F.~Salas, and E.~F. Camacho.
\newblock Input to state stability of min--max {MPC} controllers for nonlinear
  systems with bounded uncertainties.
\newblock {\em Automatica}, 42(5):797--803, 2006.

\bibitem{raimondo2009min}
D.~M. Raimondo, D.~Limon, M.~Lazar, L.~Magni, and E.~F. Ndez~Camacho.
\newblock Min-max model predictive control of nonlinear systems: A unifying
  overview on stability.
\newblock {\em European Journal of Control}, 15(1):5--21, 2009.

\bibitem{mayne2001robustifying}
D.~Q. Mayne and W.~Langson.
\newblock Robustifying model predictive control of constrained linear systems.
\newblock {\em Electronics Letters}, 37(23):1422--1423, 2001.

\bibitem{chisci2001systems}
L.~Chisci, J.~A. Rossiter, and G.~Zappa.
\newblock Systems with persistent disturbances: predictive control with
  restricted constraints.
\newblock {\em Automatica}, 37(7):1019--1028, 2001.

\bibitem{langson2004robust}
W.~Langson, I.~Chryssochoos, S.~V. Rakovi{\'c}, and D.~Q. Mayne.
\newblock Robust model predictive control using tubes.
\newblock {\em Automatica}, 40(1):125--133, 2004.

\bibitem{mayne2005robust}
D.~Q. Mayne, M.~M. Seron, and S.~V. Rakovi{\'c}.
\newblock Robust model predictive control of constrained linear systems with
  bounded disturbances.
\newblock {\em Automatica}, 41(2):219--224, 2005.

\bibitem{mayne2011tube}
D.~Q. Mayne, E.~C. Kerrigan, E.~J. Van~Wyk, and P.~Falugi.
\newblock Tube-based robust nonlinear model predictive control.
\newblock {\em International Journal of Robust and Nonlinear Control},
  21(11):1341--1353, 2011.

\bibitem{yu2013tube}
S.~Yu, C.~Maier, H.~Chen, and F.~Allg{\"o}wer.
\newblock Tube {MPC} scheme based on robust control invariant set with
  application to lipschitz nonlinear systems.
\newblock {\em Systems $\&$ Control Letters}, 62(2):194--200, 2013.

\bibitem{fleming2015robust}
J.~Fleming, B.~Kouvaritakis, and M.~Cannon.
\newblock Robust tube {MPC} for linear systems with multiplicative uncertainty.
\newblock {\em IEEE Transactions on Automatic Control}, 60(4):1087--1092, 2015.

\bibitem{marruedo2002input}
D.~L. Marruedo, T.~Alamo, and E.~F. Camacho.
\newblock Input-to-state stable {MPC} for constrained discrete-time nonlinear
  systems with bounded additive uncertainties.
\newblock In {\em Proceedings of IEEE Conference on Decision and Control},
  volume~4, pages 4619--4624, 2002.

\bibitem{richards2006robust}
A.~Richards and J.~How.
\newblock Robust stable model predictive control with constraint tightening.
\newblock In {\em Proceedings of American Control Conference}, pages
  1557--1562, 2006.

\bibitem{li2014robust}
H.~Li and Y.~Shi.
\newblock Robust distributed model predictive control of constrained
  continuous-time nonlinear systems: a robustness constraint approach.
\newblock {\em IEEE Transactions on Automatic Control}, 59(6):1673--1678, 2014.

\bibitem{sontag1995characterizations}
E.~D. Sontag and Y.~Wang.
\newblock On characterizations of the input-to-state stability property.
\newblock {\em Systems $\&$ Control Letters}, 24(5):351--359, 1995.

\bibitem{michalska1993robust}
H.~Michalska and D.~Q. Mayne.
\newblock Robust receding horizon control of constrained nonlinear systems.
\newblock {\em IEEE Transactions on Automatic Control}, 38(11):1623--1633,
  1993.

\bibitem{ICLOCS}
P.~Falugi, E.~Kerrigan, and E.~v. Wyk.
\newblock Imperial college london optimal control software (iclocs).
\newblock \url{http://www.ee.ic.ac.uk/ICLOCS/}.

\bibitem{wachter2006implementation}
A.~W{\"a}chter and L.~T. Biegler.
\newblock On the implementation of an interior-point filter line-search
  algorithm for large-scale nonlinear programming.
\newblock {\em Mathematical Programming}, 106(1):25--57, 2006.

\bibitem{mondada2009puck}
F.~Mondada, M.~Bonani, X.~Raemy, J.~Pugh, C.~Cianci, A.~Klaptocz, S.~Magnenat,
  J.~Zufferey, D.~Floreano, and A.~Martinoli.
\newblock The e-puck, a robot designed for education in engineering.
\newblock In {\em Proceedings of the 9th conference on autonomous robot systems
  and competitions}, volume~1, pages 59--65, 2009.

\end{thebibliography}

\end{document}